%% file: main.tex
\documentclass[acmsmall]{acmart}

\def\tsc#1{\csdef{#1}{\textsc{\lowercase{#1}}\xspace}}
\tsc{WGM}
\tsc{QE}
\tsc{EP}
\tsc{PMS}
\tsc{BEC}
\tsc{DE}
\usepackage{subcaption}
\usepackage{enumitem}
\usepackage{amsmath}
\usepackage{adjustbox}
\usepackage{longfigure}
\usepackage{fancyvrb}
\usepackage{makecell}
\usepackage{multirow}

\begin{document}

\title{Joint Upper \& Lower Bound Normalization for IR Evaluation}

\author{Shubhra Kanti Karmaker (``Santu'')$^*$}
\affiliation{%
  \institution{Auburn University}
  \city{Auburn}
  \state{Al}
  \country{USA}
  \postcode{36830}
}

\author{Dongji Feng$^*$}
\affiliation{%
  \institution{Auburn University}
  \city{Auburn}
  \state{Al}
  \country{USA}
  \postcode{36830}
}

\begin{abstract}
In this paper, we present a novel perspective towards IR evaluation by proposing a new family of evaluation metrics where the existing popular metrics (e.g., $nDCG$, $MAP$) are customized by introducing a query-specific \textbf{lower-bound (LB) normalization} term. While original $nDCG$, $MAP$ etc. metrics are normalized in terms of their upper bounds based on an ideal ranked list, a corresponding LB normalization for them has not yet been studied. Specifically, we introduce \textit{two} different variants of the proposed LB normalization, where the lower bound is estimated from a randomized ranking of the corresponding documents present in the evaluation set. We next conducted two case-studies by instantiating the new framework for  two popular IR evaluation metric (with two variants, e.g., $DCG^{U L}_{V_{1,2}}$ and $MSP^{U L}_{V_{1,2}}$) and then comparing against the traditional metric without the proposed LB normalization. Experiments on two different data-sets with eight Learning-to-Rank (LETOR) methods demonstrate the following properties of the new LB normalized  metric: 
1) Statistically significant differences (between two methods) in terms of original metric no longer remain statistically significant in terms of Upper Lower (UL) Bound  normalized version  and vice-versa, especially for \textit{uninformative} query-sets. 2) When compared against the original metric, our proposed UL normalized metrics demonstrate higher \textit{Discriminatory Power} and better \textit{Consistency} across different data-sets.
These findings suggest that the IR community should consider UL normalization seriously when computing \textit{nDCG} and \textit{MAP}  and more in-depth study of UL normalization for general IR evaluation is warranted.

% 2) Although original metric scores drop once we apply LB normalization, percentage differences among multiple LETOR method-pairs actually increase (more distinguishable); 3) For a number of closely performing LETOR method-pairs, statistically significant differences (between two methods) in terms of original  metric no longer remain statistically significant in terms of LB normalized  version and vice-versa; 4) This change in statistical significance is more prominent in case of \textit{broad} (exploratory) queries compared to \textit{focused} ones for all three metrics on both data-sets. These findings suggest that the IR community should consider LB normalization seriously when computing \textit{nDCG}, \textit{ERR} and \textit{MAP} and more in-depth study of LB normalization for general IR evaluation is warranted.
\end{abstract}




\maketitle
\def\thefootnote{*}\footnotetext{These authors contributed equally to this work}\def\thefootnote{\arabic{footnote}}

\input{sections/0Introduction}

\input{sections/1Related}
\input{sections/2nDCG}

\input{sections/3General}

\input{sections/4ExperimentalDesign}

\input{sections/5RRnDCG}

\input{sections/7MAP}

\input{sections/8Conclusions}
\bibliographystyle{cas-model2-names}

\bibliography{refs}

\end{document}

%% file: sections/0Introduction.tex
\vspace{8mm}
\section{Introduction}\label{sec:intro}
Empirical evaluation is a key challenge for any information retrieval (IR) system. The success of an IR system largely depends on the user's satisfaction, thus an accurate evaluation metric is crucial for measuring the perceived utility of a retrieval system by the real users. While original $nDCG$~\cite{jarvelin2002cumulated}, $MAP$~\cite{caragea2009mean} etc. metrics are normalized in terms of their query-specific upper bounds based on an ideal ranked list, a corresponding query-specific LB normalization for them has not yet been studied. For instance, the normalization term in \textit{nDCG} computation is the \textit{Ideal DCG} at cut-off $k$, which converts the metric into the range between 0 and 1. On the other hand, $MAP$ is normalized by the maximum possible \textit{Sum of Precision} (SP) scores at cut-off $k$. Thus, \textit{Ideal DCG} and \textit{Sum of Precision} (SP) scores essentially serve as the query-specific upper-bound normalization factor for metric \textit{nDCG} and \textit{MAP}, respectively. 

Interestingly, above two popular metrics do not include a similar query-specific lower-bound (LB) normalization factor (the current widely used assumption for lower-bound is \textbf{zero} across all queries). However, each query is different in terms of its difficulty (informative/ uninformative/ distractive), user's intent (exploratory/ navigational), distribution of relevance labels of its associated documents (hard/ easy) and user's perceived utility at different cut-off $k$, essentially implying different low-bounds for each of them. Therefore, an accurate estimation of an evaluation metric should not only involve an upper-bound normalization (e.g., Ideal DCG, SP etc.), but also a proper query-specific lower-bound normalization.

Consider the case of re-ranking where an initial filtering has already been performed given a query and as expected, a large number of associated documents in the filtered set are highly relevant. In this case, even just a random ranking of those documents will yield a high accuracy as most of the documents are highly relevant anyway. This means that even if a ranker does not learn anything meaningful and merely ranks documents randomly, it can still achieve a very high score in terms of the original metric. In other words, the \textit{expected} value/lower-bound of the original metric in this case is very high because of the skewed relevance label distribution of the associated documents and this factor should be accounted for while measuring the ranker's quality. In summary, a proper lower-bound normalization is essential for IR evaluation metrics to accurately measure the quality of a ranker as well as for a fairer comparison across multiple ranking methods.

What does query-specific lower-bound normalization mean for an IR evaluation metric? How can we come up with a more realistic lower-bound for each query and include it with the original IR  metric computation? 
One way to address this issue is to introduce a penalty term inside the formula of different IR evaluation metrics which will penalize queries with high expected values of the same metric. In other words, given a query, we propose to use the expected value of the particular evaluation metric as a query-specific lower-bound of the same metric for that query, which can yield customized lower-bounds for different queries and thus, ensure fairer treatment across all queries with different difficulty levels.

With the observation that both \textit{nDCG} and \textit{MAP} metrics only involve query-specific upper-bound normalization (e.g., normalization with ideal DCG for \textit{nDCG} computation, while MAP is normalized by the maximum possible \textit{Sum of Precision}); none of them include a query-specific lower-bound normalization. 
In this paper, we proposed a new general framework for IR evaluation with both upper and lower bound normalization and instantiated the new framework for two popular IR evaluation metric: \textit{nDCG} and \textit{MAP} by computing a more reasonable(non-zero) lower-bound, Specifically, we introduce two different variants of the framework, i.e., $V_1$, $V_2$, which are essentially two different ways to introduce penalty in terms of normalization with a query-specific upper and lower-bound of the metric (see section~\ref{sec:general} for more details). We then show how we can compute a more realistic query-specific lower-bound for the two metrics by computing its expected value for each query in case of a randomized ranking of the corresponding documents, and then, use this lower-bound as a penalty term while computing the new metric. \textit{The intuition here is that an intelligent ranking method should perform at least as good as a random-ranking algorithm, which naturally inspired us to use the expected value in case of random ranking as our lower-bound}. Finally, for each metric we also theoretically prove the correctness the expected lower-bound (Derivation details can be found in each case-study sections).

% To answer these questions, we instantiated the new framework for the popular \textit{nDCG} metric by introducing a penalty term inside the original metric which will penalize queries with high expected values of the same metric. In other words, given a query, we propose to use the expected value of $nDCG$ as a query-specific lower-bound of the same metric for that query, which can yield customized lower-bounds for different queries.

% To elaborate, we conducted a case-study in this paper by proposing an extension of the original \textit{nDCG} metric, i.e., $DCG^{U L}$,  which incorporates a query-specific lower-bound normalization term (see section~\ref{sec:general} for more details). We then show how we can compute a more realistic query-specific lower-bound for \textit{nDCG} by computing its expected value for each query in case of a randomized ranking of the corresponding documents, and then, use this lower-bound  as a penalty term. \textit{The intuition here is that an intelligent ranking method should at least perform as good as a random-ranking method}. This naturally inspired us to use the expected value in case of random ranking as the lower-bound.

Next, we investigated the implications of upper lower-bound normalization on the original IR metric. How it may impact IR evaluation in general and more importantly, which metric is better? Why should we care? To answer these questions, we have conducted extensive experiments on two popular Learning-to-Rank (LETOR) data-sets with eight LETOR methods including RankNet~\cite{burges2005learning},  RankBoost~\cite{freund2003efficient}, AdaRank~\cite{xu2007adarank},  Random Forest~\cite{breiman2001random}, LambdaMART~\cite{burges2010ranknet},  CoordinateAscent~\cite{metzler2007linear}, ListNet~\cite{cao2007learning} and L2 regularized Logistic Regression~\cite{fan2008liblinear,lin2008trust}. Experimental results demonstrate that a significant portion of the queries in popular benchmark data-sets produced a high LB normalization factor, verifying that LB normalization can indeed alter the relative ranking of multiple competing methods (confirmed by Kendall's $\tau$ tests~\cite{sakai2016simple,sakai2006evaluating}) and thus, should not be ignored. At the same time, for a number of closely performing LETOR method-pairs, statistically significant differences in terms of original metric no longer remain statistically significant in terms of LB normalized metric and vice-versa, especially for \textit{uninformative} query-sets (see section~\ref{sec:motivation} for a concrete definition), suggesting LB normalization yields different conclusions than the original metric.

Next, we compare original metric against Upper lower bound normalized version from two perspectives: \textit{Distinguishability} and  \textit{Consistency}. In case of discriminative power, we followed Sakai~\cite{sakai2006evaluating,sakai2011using} to use student's t-test as well as computed ``Percentage Absolute Differences''  to quantify distinguishability and found that UL bound normalized version can better distinguish  between two closely performing LETOR methods in case of \textit{uninformative} queries. For consistency, we performed swap rate tests and and found that for $MSP^{U L}$ provide a better performance in terms of \textit{Consistency} while  $DCG^{U L}$ does not compromise in terms of \textit{Consistency}. 

These findings suggest that the community should rethink about IR evaluation and consider LB normalization seriously. In summary, we make the following contributions in the paper:

\begin{enumerate}[leftmargin=*,itemsep=0ex,partopsep=0ex,parsep=0ex]

\item  We propose an extension of traditional IR evaluation metrics which includes a lower bound (LB) normalization term, and systematically perform two case-studies by showing how LB normalization can be materialized for \textit{nDCG} and \textit{MAP}.

\item We propose two different variants of the proposed upper lower-bound normalized version for two popular IR evaluation metrics.

\item  We show how we can compute a more realistic query-specific lower-bound for two IR evaluation metrics by computing its expected value for each query in case of a randomized ranking of the document collection and also theoretically prove its correctness.

\item We conducted extensive experiments to understand the implications of LB normalized metric and compared our proposed metric against the original metric from two important perspectives: \textit{Distinguishability} and  \textit{Consistency}.

\item Our proposed framework is very general and can be easily extended to other IR evaluation metrics or evaluation metrics in other domain.
\end{enumerate}

The rest of the paper is organized as follows: Section \ref{sec:related} reviews related works from the past literature. Section \ref{sec:nDCG} provides essential background about our two experimental metric computation and motivation for lower-bound normalization. In section \ref{sec:general}, we first present the framework with query-specific upper and lower bound normalization. Section~\ref{sec:case4nDCG_ERR_MAP} presents the experiment details and results. Finally, section~\ref{sec:conclusion} concludes our paper with discussions and possible future directions.

%% file: sections/1Related.tex
\section{Related Work}\label{sec:related}
% \noindent\textbf{DISCARD , no Fairness in IR:} The definition of fairness and related concerns have been widely discussed in previous research and becoming an important topic in the community of IR and AI contexts. \cite{lipani2016fairness} focused on the fairness of cornerstones in IR evaluation research such as \textit{precision} and \textit{recall}. Meanwhile, \cite{tonon2015pooling} introduced a measure called \textit{Fairness Score} to measure the fairness of judgement poll.
% On the other hand, the research about the fairness of algorithms were also been explored: \cite{gao2019fair} considered the fairness as an optimization problem and proposed a framework that is able to offer a novel perspective into the optimization with fairness constraints problems.  \cite{pathiyan2021evaluating} focused on the fairness of argument retrieval and analyzed a range of non-stochastic fairness-aware ranking and diversity metrics to evaluate the extent to which argument stances are fairly exposed in argument retrieval system. 

% \noindent\textbf{Multi-stage retrieval systems:} Multi-stage retrieval systems consisted of a candidate generation process followed by one or more re-ranking stages as in practice it is difficult to directly apply machine-learned models over the \textit{entire} collection~\cite{asadi2013effectiveness,lin2021multi}. The scenario of our work is to evaluate the \textit{re-ranking} performance of LETOR methods w.r.t. a candidate list of potentially-relevant documents.

\noindent\textbf{Traditional IR evaluation metric:} Many metrics have been introduced for IR system evaluation~\cite{manning2008introduction,kanoulas2015short} in recent years. Two most frequent and basic metrics for the performance evaluation of IR system are \textit{precision} and \textit{recall}, especially for extraction tasks~\cite{karmaker2016generative,sarkar2022concept}. Empirical studies of retrieval performance have shown a tendency for \textit{precision} to decline as \textit{recall} increases~\cite{buckland1994relationship}. Due to the trade-off between the two basic calculations, researchers also use other complex single metrics such as \textit{F-measure} which can evenly weight the \textit{precision} and \textit{recall}. Other popular metrics such as $MAP$ (Mean Average Precision) and Normalized Discounted Cumulative Gain ($nDCG$)  are also widely used as offline evaluation standards. Different metrics have different hyper-parameters for users to choose based their own preferences.

% Meanwhile, \textit{AP} (Average Precision) and \textit{MAP} (Mean Average Precision) are also widely used as offline metrics.  

\smallskip
\noindent\textbf{nDCG:} \textit{nDCG} is the normalized version of \textbf{D}iscounted \textbf{C}umulative \textbf{G}ain (\textit{DCG}), where the normalization term is essentially a \textit{query-specific} upper-bound (i.e., normalization with \textit{Ideal DCG}), which converts the metric into the range between 0 and 1 ~\cite{jarvelin2002cumulated}. It has become one of the most important metrics because it can be applied to multi-level relevance judgments and is sensitive to small changes in a ranked list, and it has become the most popular measure for evaluating Web search and Learning-to-Rank algorithms~\cite{valizadegan2009learning}. 
Many researchers have investigated its properties (see, e.g., ~\cite{yilmaz2008simple, ravikumar2011ndcg, wang2013theoretical}). 
The fact that the general concept of $nDCG$ can be implemented in a variety of ways was recognized in the previous work~\cite{kanoulas2009empirical},  where the authors scrutinized how to choose from a variety of discounting functions and different ways of designing the gain function to optimize the efficiency or stability of $nDCG$~\cite{karmaker2017application}. Previous research has also shown that with different gain functions, $nDCG$ may lead to different results and the discounting coefficients do make a difference in evaluation results as compared to using uniform weights ~\cite{voorhees2001evaluation}. Regarding $nDCG$ cutoff-depths, Sakai and others ~\cite{sakai2007reliability} have researched the reliability of $nDCG$ by establishing that it is highly correlated with average precision if the cutoff-depth $k$ is big enough. According to a recent research~\cite{DBLP:conf/cikm/KarmakerSZ20}, conventional $nDCG$ score results in significant variance in response to the $k$ value and urged for query-specific customization of $nDCG$ to acquire more trustworthy conclusions. Additionally, Lukas et.al~\cite{gienapp2020estimating} proposed a measure to explicitly reflect a system's divergence by comparing the query-level $nDCG$ with a randomized ranked $nDCG$, which they called $RNDCG$. They claimed that this measure can capture the general trend of query difficulty by the ratio-based score and further improving several issues such as selecting a specific set of query~\cite{gienapp2020estimating}.

% \noindent\textbf{ERR:} 
% One disadvantage of nDCG is the basic assumption that the usefulness of the document at rank $i$ is independent of the usefulness of the documents at rank less than $i$.~\cite{shi2013xclimf}. 
% To address this issue, ~\cite{chapelle2009expected} suggested the
% Expected Reciprocal Rank (ERR) metric based on a cascade-style user browsing model. It was developed as a generalization of $earlier$ $reciprocal$ $rank$ (RR) metric, in response to the emergence of $graded$ $relevance$ $judgements$ ~\cite{jarvelin2002cumulated}.
% In their research, they demonstrated that cascade model captures real user browsing behavior better and ERR correlated with a range of click-based metrics much better than other editorial metrics such as DCG. Many researchers have proposed approaches for adjusting parameters of ERR model. For instance, ~\cite{logachev2012tuning} argued that these parameters of ERR model should be adjusted more accurately and set by analyzing real user's behaviour therefore they proposed two approaches for adjusting parameters and observed that their parameters are largely different from the commonly used parameters of ERR model.
% Yury and others~\cite{logachev2012optimizing} optimized the parameters by maximizing Person weighted correlation between ERR and several online click metrics.
% Previous researches also utilize ERR for different tasks. ~\cite{shi2013xclimf} discovered the effectiveness of optimizing ERR for data with multiple levels of relevance(i.e.,xCLiMF model) and found outperforms on a few baseline methods.

\smallskip
\noindent\textbf{MAP:}
Average precision (AP) is one of most commonly used indicator for evaluating ranked output in IR experiments for a number of reasons as it already known to be stable~\cite{buckley2017evaluating} and and highly informative measure~\cite{aslam2005maximum}. Whereas
Mean Average Precision (MAP) ~\cite{caragea2009mean} is the average of AP of each class which can reflect the overall performance among multiple topics.
However, the main criticism to MAP is that it is based on the assumption that retrieved documents can be considered as either relevant or non-relevant to user's information need, which is not accurate. Previous researchers have studied the properties of MAP in terms of  different relevance judgement. Yilmaz et.al ~\cite{yilmaz2006estimating}, for instance, proposed  different variant of AP for addressing incomplete and imperfect relevance judgements, where they consider the document collection is dynamic, as in the case of web retrieval, and they use a expectation of randomly sample from the depth-100 pool. Furthermore, ~\cite{robertson2010extending} proposed a extended Average Precision called Graded Average Precision (GAP) which can tackle the cases of multi-graded relevance.

% \noindent\textbf{Query Specific Customization of nDCG:} 
% Recently, Karmaker et.al.~\cite{DBLP:conf/cikm/KarmakerSZ20} conducted an in-depth study with Learning-to-Rank (LETOR) methods on the implications of using \textit{query-specific} cut-off while computing $nDCG$, particularly to see whether using a customized cut-off $k$ for each individual query would lead to a different conclusion about the relative performances of multiple LETOR methods than using the conventional query-independent $nDCG$ would otherwise. Their initial results show that the relative ranking of LETOR methods using \textit{query-specific} cut-off $k$ can be dramatically different (at the individual query level) from those using $nDCG$ with constant cut-off $k$. In other words: \textit{For a particular query, while method $X$ is better than method $Y$ in terms of $nDCG@k_1$, it is possible that method $Y$ is better than method $X$ in terms of $nDCG@k_2$, where $k_1\ne k_2$.} In fact, they found many such cases in their study, specifically, $959$ cases within a set of $1000$ queries. Their study confirms that an appropriate query-specific customization of $nDCG@k$  is important for accurate evaluation of a ranking system. 

\smallskip
\noindent\textbf{Query Specific Customization for General IR Evaluation:} Previous work has explored how to incorporate query specific customization for IR evaluation metrics in general. For example, Moffat et.al.~\cite{moffat2013users} followed by Bailey et.al.~\cite{bailey2015user} argued that user behavior varies on a per-topic basis depending on the nature of the underlying information need, and hence that it is natural to expect that evaluation parameterization should also be variable. Billerbeck et.al. studied the optimal number of top-ranked documents that should be used for extraction of terms for expanding a query~\cite{billerbeck2004questioning}. Such work has shown the need to adapt a ranking function to each individual query. Egghe et.al.~\cite{egghe2008measures} demonstrated precision, recall, fallout and miss as a function of the number of retrieved documents and their mutual interrelations. Kuzi et.al.~\cite{kuzi2019analysis} presented a Best-Feature Calibration (BFC) strategy for analyzing learning to rank models and used this strategy to examine the benefit of query-level adaptive training, which demonstrated the importance of query-specific parameters in IR evaluation once again.

\smallskip
\noindent\textbf{IR Evaluation with Variable Parameterization:} Query specific customization can be viewed as a special case of variable parameterization for IR evaluation metrics, which has been explored previously. Webber et.al.~\cite{webber2010effect} explored the role that the metric evaluation depth $k$ plays in affecting metric
values and system-versus-system performances for two popular families of IR evaluation metrics: i.e., recall-based and utility-based metrics. Study by Jiang et.al.~\cite{jiang2016adaptive} showed that the adaptive effort metrics can better indicate user's search experience compared with conventional metrics. Yilmaz et al. showed users are more likely to click on relevant results~\cite{yilmaz2010expected} and also examined the differences between searcher's effort (dwell time) and assessor's effort (judging time) on results, and features predicting such effort~\cite{yilmaz2014relevance}. Sakai et.al.~\cite{sakai2008modelling} modeled a user population  to assess the appropriateness of different evaluation metrics.

\smallskip
\noindent\textbf{Distinction from prior work:} Our work completely differs from the previous effort as our goal is to investigate the impact of lower-bound normalization on the prominent evaluation metrics. To the best of our knowledge, there has never been a systematic study of query-specific lower-bound normalization for IR evaluation metrics. Furthermore, our work is groundbreaking in that it proposes a generic upper and lower-bound (UL) normalization framework and effectively applies it to two prominent evaluation metrics. We additionally compute an expectation over a randomized ranked list to estimate a more realistic lower-bound and also give the derivation. Our research clearly articulates the effects of such lower-bound normalization on two popular evaluation metrics and lays the foundation for future research in this direction.

%% file: sections/2nDCG.tex
\section{Revisiting Original Metrics}\label{sec:nDCG}
In this section, we provide some essential background about $nDCG$ and $MAP$ computation and also provide our motivation of lower-bound normalization for the two metrics.

\subsection{Computation of the standard nDCG}
The principle behind  Normalized Discounted Cumulative Gain ($nDCG$) is that documents appearing lower in a search result list should contribute less than similarly relevant documents that appear higher in the results~\cite{jarvelin2002cumulated}. This is accomplished by introducing a penalty term that penalizes the gain value logarithmically proportional to the position of the result~\cite{wang2013theoretical}. Mathematically:

\begin{equation}
{ DCG@k}\ =\ \sum_{i=1}^{k}\frac{2^{R_i}\ -\ 1}{\log_b(i+1)}
\end{equation}

Here, i denotes the position of a document in the search ranked list and ${R_i}$ is the  relevance label of the $i-th$ document in the list, cutoff $k$ means $DCG$ accumulated at a particular rank position $k$, the discounting coefficient is to use a log based discounting factor $b$ to unevenly penalize each position of the search result. $nDCG@k$ is $DCG@k$ divided by maximum achievable $DCG@k$, also called Ideal $DCG$(\textit{IDCG@k}), which is computed from the ideal ranking of the documents with respect to the query.

\begin{equation}
nDCG@k\ =\ \frac{DCG@k}{IDCG@k}
\end{equation}

\subsection{Computation of the standard MAP}
For our second case study, we selected another popular evaluation metric called Mean Average Precision ($MAP$). In the field of information retrieval, precision is the fraction of retrieved documents that are relevant to the query. The formula is given by: $Prec = {TP}/{(TP + FP)}$, where, $TP$ and $FP$ stands for \textit{True Positive} and \textit{False Positive}, respectively. Precision at cutoff $k$ is the precision calculated by only considering the subset of retrieved documents from rank $1$ through $k$. However, the original precision metric is not sensitive to the relative order of the ranked documents, hence, we do not consider it for our exploration.

%and ($TP$) is the item that the test makes a positive prediction, and the subject has a positive result under the gold standard, False Positive ($FP$) is the item that the test makes a positive prediction however has a negative result under the gold standard. 

A related popular metric, which is sensitive to the relative order of the ranked documents, is \textit{Average Precision}, which computes the sum of precision scores at each rank where the corresponding retrieved document is relevant to the query. 

\vspace{-1mm}
\begin{equation} \label{equ:AP@k}
    AP@k = \frac{1}{k} \sum_{i=1}^{k} Prec(i) \cdot R_i
\end{equation}

Here, $R_i$ is an indicator variable that says whether $i^{th}$ item is relevant ($R_i = 1$) or non-relevant ($R_i = 0$). From Formula \ref{equ:AP@k}, we can see $AP@k$ is already normalized by the the maximum possible \textit{Sum of Precision} (SP), which is $k$ in this case by assuming a precision value of $1.0$ for every position from $1$ to $k$. Thus, $AP@k$ is already upper-bound normalized version of $SP@k$, like $nDCG@k$ is for $DCG@k$. Finally, Mean Average Precision ($MAP$) of a set of queries is defined by the following formula, where, $|Q|$ is the number of queries in the set and $AP(q)$ is the average precision ($AP$) for a given query $q$. 

\vspace{-3mm}
\begin{equation*}
    MAP = \frac{ \sum_{q=1}^{|Q|} AP(q) } {|Q|} 
\end{equation*}

In summary, $AP$ is essentially an upper-bound normalized version of \textit{Sum of Precision} ($SP$), which is defined as follows:

\noindent{\bf Sum of Precision (SP)}:  \textit{SP computes the summation of the precision scores at all ranks (from 1 to rank $k$), where the retrieved document is relevant to the query without any upper or lower bound normalization.}  

\begin{equation}
    SP@k = \sum_{i=1}^{k}Prec(i) \cdot R_{i}
\end{equation}

% Below, we will present how we can compute a realistic lower bound for \textit{Sum Precision} ($SP$) by computing its expected value in case of a randomly ranked list of documents.

\subsection{Motivation for Lower-bound Normalization}\label{sec:motivation}

A closer look into the formula of conventional \textit{nDCG} and \textit{MAP} shows that the two metrics incorporate only a query-specific upper-bound normalization (i.e.,\textit{IDCG} is actually an upper-bound normalization term).
However, as mentioned in section~\ref{sec:intro}, each query is different in terms of difficulty (hard/ easy), informativeness (informative/ uninformative/ distractive), user's intent (exploratory/ navigational); as such, they have different expected value for the lower-bound of different evaluation metric. Thus, an accurate estimation of average $nDCG$ and  $MAP$  should include different lower-bounds for different queries.

The main motivation of our work is to relax the incorrect assumption of uniform lower-bound (of $nDCG$ and $MAP$) across all queries while evaluating IR systems. We propose that an accurate evaluation metric should customize for each query and normalize with respect to both query-specific upper and lower-bound. A follow-up question that arises immediately is the following: How can we estimate a realistic lower-bound of an IR evaluation metric? While original implementation of above two metric assume \textit{zero} as the lower bound, previous work proposed to use worst possible ranking score as the lower bound~\cite{cikmnDCGMinMaxNorm} to achieve a standardized range, we argue that this lower bound can be further constrained by using the score of a randomly ranked list for each query. \textit{The justification behind this choice is that a reasonable ranking function should be at least as good as the method that ranks documents merely randomly and should be penalized in cases where it performs worse than random}. 

\begin{figure}[!htb]
    \centering
    \includegraphics[width=0.8\linewidth]{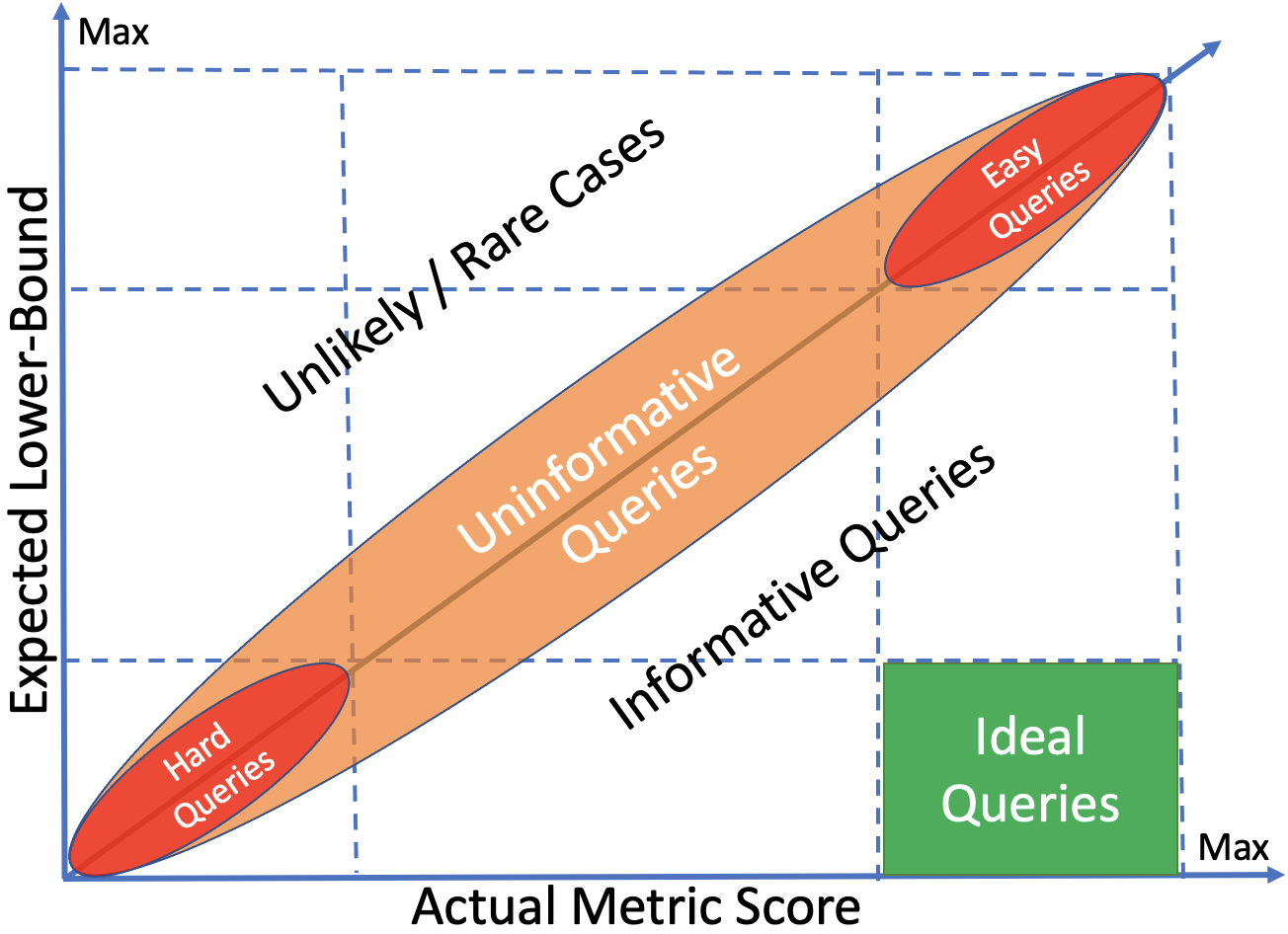}
    \vspace{-2mm}
    \caption{Query types with different Lower Bounds of evaluation metric.}
    \label{fig:query_categorization}
    \vspace{-2mm}
\end{figure}

To better motivate LB normalization, we first define the following types of queries, which we will use throughout the rest of the paper:

\begin{enumerate}[leftmargin=*,itemsep=0ex,partopsep=1ex,parsep=0ex]
    \item {\textbf{Informative Queries:}} These are queries where a \textit{reasonable} ranking method performs significantly better than a pure random ranking system. Essentially, these are queries which contain the ``right'' keywords to find out the most relevant documents according to the user's information need. Therefore, the actual evaluation metric scores are much higher than the expected lower-bound (the lower triangle region of the plot \ref{fig:query_categorization}).
    
    \smallskip
    {\textbf{Ideal Queries:}} These are special cases of \textit{Informative}
    queries where the difference between actual evaluation metric score and random ranked metric score (lower-bound) is the largest.

    \item {\textbf{Uninformative Queries:}} These are queries where a \textit{reasonable} ranking method performs close to a pure random ranking system. In other words, these are queries which does not offer much value in finding out the most relevant documents. Therefore, the actual evaluation metric scores are similar to the expected lower-bound (region around the diagonal line). There are two special cases for Uninformative queries as defined below:

    \begin{enumerate}[leftmargin=*,itemsep=0ex,partopsep=1ex,parsep=0ex]
        \item {\textbf{Hard Queries}:} Hard queries are special cases of \textit{Uninformative} queries, where both \textit{reasonable} ranking methods as well as pure random ranking systems demonstrate poor performance. This usually happens in cases where there is no/very few relevant documents in the entire corpus.
        
        \item {\textbf{Easy Queries:}} Easy queries are special cases of \textit{Uninformative} queries, where both \textit{reasonable} ranking methods as well as pure random ranking systems demonstrate very high performance. This usually happens in cases where there is a lot of relevant documents in the corpus (for example, in case of re-ranking in multi-stage ranking systems\cite{asadi2013effectiveness, clarke2016assessing,tonellotto2013efficient}) and there is little room for improving beyond random ranking.
    \end{enumerate}

\end{enumerate}

Figure~\ref{fig:query_categorization} shows an illustration of different types of queries with different combinations of lower-bound evaluation metric and actual metric score. As apparent from Figure~\ref{fig:query_categorization}, the proposed LB normalization is expected to have large penalty on uninformative queries including special cases like hard queries (lack of relevant document scenarios) and easy queries (re-ranking scenarios). On the other hand, LB normalization will have minimal impact in case of \textit{Ideal} queries as the lower-bound tends to zero and actual metric score is very high. However, as demonstrated by our experiments, real-world queries are not \textit{Ideal} always and hence, a proper LB normalization is necessary while computing $nDCG$ and  $MAP$  scores because: $1)$ It better captures the difficulty as well as variations across different queries. $2)$ It makes comparison and averaging across different queries fairer.

%% file: sections/3General.tex
\section{IR Evaluation with Joint Upper \& Lower Bound Normalization}\label{sec:general}
% \Dongji{general just the introduce of framework and dataset, put case studies as another section })

Assume that $A@k$ is the standard evaluation metric and $k$ is the cutoff rank. Before introducing the generic IR evaluation framework with both upper \& lower bound (UL) normalization, we first define the following terms.

\begin{itemize}
    \item {{$\bf IUB[A@k]$:}} Given a particular query and an associated collection of documents (each with a distinct relevance labels), $IUB[A@k]$ (\textbf{I}deal \textbf{U}pper \textbf{B}ound for $A@k$) is the value that $A@k$ assumes in case of \textit{\textbf{perfect} ranking} of the document collection.

    \item {\textbf{$\bf RLB[A@k]$:}} Given a particular query and an associated collection of documents (each with a distinct relevance labels), $RLB[A@k]$ (\textbf{R}andomized \textbf{L}ower \textbf{B}ound for $A@k$) is the value that $A@k$ assumes in case of \textit{\textbf{random}} ranking ($E[A@k]$) of the document collection.
    
    \item{\textbf{Upper-Bound Normalization}: } Given a particular query and an evaluation metric $A@k$, Upper-bound normalization of the metric is defined as $[A@k]^{U}=\frac{A@k}{IUB[A@k]}$.
\end{itemize}

Now, we introduce two different variations of Joint Upper \& Lower Bound Normalization, which is denoted by, $[A@k]^{U L}$. We call the two versions as $V_1$, $V_2$.

\begin{equation}\label{equ:V1}
    [A@k]^{U L}_{V_1} = \left(\frac{A@k}{IUB[A@k]}\right)\left(\frac{A@k}{(A@k + RLB[A@k])}\right)
\end{equation}
\vspace{2.25mm}

\begin{equation}\label{equ:V2}
   [A@k]^{U L}_{V_2}= 
\begin{cases}
     \frac{A@k - RLB[A@k]}{IUB[A@k] - RLB[A@k]},& \text{if } A\geq RLB\\\\
     
    \frac{A@k - RLB[A@k]}{RLB[A@k]},              & \text{otherwise}
\end{cases}
\end{equation}

\vspace{0.7mm}

% \begin{equation}\label{equ:V3}
% [A@k]^{U L}_{V_3} =  \ \frac{abs(A@k - RLB[A@k])}{IUB[A@k] - RLB[A@k] - min(0, A@k - RLB[A@k])}
% \end{equation}
% \vspace{0.25mm}

In the first Equation\textbf{~\ref{equ:V1}}, we introduce a linear penalty term for Upper Lower Bound Normalization while in the second Equation\textbf{~\ref{equ:V2}} we introduce a non-linear penalty term. 
The intuition of above two Equation is that we want to penalize methods for queries where it performs close to a random ranking method, i.e., the difference between $A@k$ and $RLB[A@k]$ is minimal (the uninformative queries): $|A@k - A[A@k]| \equiv 0$. Even if a ranker achieves high $A@k$ in this case, it does not necessarily mean it is an ``intelligent'' ranker as the ``vanilla'' random ranking method can achieve similar performance as well. So, the reward for the method in this case should be discounted. Therefore, to truly distinguish between an ``intelligent'' and ``vanilla'' ranking method, it is important to penalize the traditional metric with a more realistic lower-bound, e.g., score w.r.t. a randomly ranked collection. In other words, for a ranking algorithm to claim a high $A@k$ score, it must perform significantly better than the random ranking baseline.
% The main intuition behind these two versions is that we want to penalize queries with a higher Randomized Lower Bound ($RLB$).
% Imagine an extreme case that a query with $100$ documents in the collection and $90$ of which are highly relevant; in this case, if we just randomly rank those documents, the evaluation metric will still yield a high score even though no intelligent processing was done. In other words, traditional metrics will yield high numbers for any ``dumb'' ranking method due to the higher $RLB$ of the metric. Thus, to truly distinguish between an ``intelligent'' and ``dumb'' ranking method, it is important to penalize the traditional metrics with a more realistic lower-bound score w.r.t. a randomly ranked collection. In other words, for a ranking algorithm to claim a high accuracy number, it must outperform the random ranking baseline considerably.

\subsubsection{\bf{Range of LB normalized Metric:}}\label{sec:rangeOfLB} 
It should be noted that $V_1$ and $V_2$ are just two different ways to introduce the penalty for higher $RLB$ and obviously, more variants are possible while the basic idea remains the same. As can be seen from Equation\textbf{~\ref{equ:V1}}, $V_1$ includes an additional multiplicative term that penalizes the original metric with the $RLB$ term in the denominator and the range of the metric is still bounded between $0$ and $1$. $V_2$  (Equation\textbf{~\ref{equ:V2}}) works as follows: instead of range $[0,1]$, it extends the range from negative to positive real numbers yielding negative numbers for a ranking method which performs worse than the random ranking baseline.
\textbf{In summary}, for Equation\textbf{~\ref{equ:V1}}, the range is still $[0,1]$; while for Equation\textbf{~\ref{equ:V2}}, the range of the metric is extended from $-1$ to $+1$ where, $+1$ means perfect ranking, $0$ means randomized ranking and $-1$ means all irrelevant results.

% It should be noted that $V_1$, $V_2$ are just two different ways to introduce the penalty for higher $RLB$ and obviously, more variants are possible while the basic idea remains the same. As can be seen from equation~\ref{equ:V1}, $V_1$ includes an additional multiplicative term that penalizes the original metric with the $RLB$ term in the denominator and the range of the metric is still bounded between $0$ and $1$. $V_2$  (equation~\ref{equ:V2}) works as follows: instead of considering the lower bound as $0$, it extends the range from negative to positive real numbers yielding negative numbers for a ranking method which performs worse than the random ranking baseline (e.g., use $0$ as random baseline and $-1$ as lower bound ).

%% file: sections/4ExperimentalDesign.tex
\subsection{Data Set}
\vspace{-1mm}
We used two popular LETOR data-sets, i.e., ``MSLR-WEB30K''~\cite{qin2010microsoft} and "MQ2007"~\cite{DBLP:journals/corr/QinL13} for our experiments. The first and second data-set includes 30,000 and 1,700 queries respectively and have widely been used as benchmarks for LETOR tasks~\cite{ganjisaffar2011bagging,shukla2012parallelizing,jia2021pairrank}. In these data-sets, each row corresponds to a query-document pair. The first column represents the relevance label of the pair, the second column is the query id, and the rest of columns represent features. The relevance scores are represented by an integer scale between 0 to 4 for ``MSLR-WEB30K'' and between 0 to 2 for ``MQ2007'', where 0 means non-relevant and 4(2) means highly relevant. The larger the value of relevance label, the more relevant the query-document pair is. Features related to each query-document pair is represented by a 136 dimensional feature vector for ``MSLR-WEB30K'' and 46 dimensional feature vector for ``MQ2007'' data-set ~\cite{DBLP:conf/cikm/KarmakerSZ20}. For more details on how the features were constructed, see ~\cite{DBLP:journals/corr/QinL13} and ~\cite{qin2010microsoft}.

We randomly sampled 10,000 queries from the ``MSLR-WEB30K'' and 1000 queries from ``MQ2007'' individually. For ``MSLR-WEB30K'', the average number of documents associated with each query was 119.06; while for ``MQ2007'', the number was 41.47. We kept all the features available (136 for ``MSLR-WEB30K'' and 46 for ``MQ2007'') for all experiments conducted in this paper.

\begin{table}[h]
 	\begin{tabular}{r|c||r|c}\hline
 	{\bf Algorithm} &{\bf Short form }  & {\bf Algorithm} &{\bf Short form }\\\hline\hline
	RankNet~\cite{burges2005learning} & RNet & LambdaMART~\cite{burges2010ranknet} & LMART\\\hline
	
 	RankBoost~\cite{freund2003efficient} & RBoost & CoordinateAscent~\cite{metzler2007linear}& CA\\\hline
 	
 	AdaRank~\cite{xu2007adarank}& ARank & ListNet~\cite{cao2007learning}& LNet\\\hline

 	Random Forest~\cite{breiman2001random} & RF & Logistic Regression~\cite{fan2008liblinear} & L2LR\\\hline
  	\end{tabular}
\caption{Popular learning to rank algorithms}
\vspace{-2mm}
\label{table:algorithms}
\end{table} 

\subsection{Learning to Rank (LETOR) Methods} 
Table \ref{table:algorithms} contains eight prominent LETOR approaches along with popular classification and regression methods used for ranking applications. We also assign acronyms to each approach for notational convenience, which we will use throughout the rest of the paper.

%% file: sections/5RRnDCG.tex
\subsection{Case Study 1: \textit{nDCG} with Joint Upper \& Lower Bound Normalization} \label{sec:case4nDCG_ERR_MAP}
In each case study sections we first describe how to compute a more realistic lower-bound for the corresponding metric, ($nDCG$ for the first case study) i.e., the expected $nDCG$ in case of random ranking. 
Although Lukas et.al~\cite{gienapp2020estimating} proposed to use the expectation to estimate this value, no derivation process provided.
Note that, $nDCG$ is already an upper-bound normalized version of $DCG$. Therefore, we start with the original metric $DCG@k$, where, $RLB[DCG@k]$ is the expected $DCG@k$ computed based on a randomly ranked list. Thus, we use the terms $E[DCG@k]$ and $RLB[DCG@k]$ interchangeably throughout the paper. Additionally, LB-normalized \textit{nDCG} and upper lower bound(UL) normalized \textit{DCG} also mean the same thing and we will use them interchangeably throughout the paper as well.

\subsubsection{\textbf{Expected DCG@k: }}\label{sec:Expected4nDCG} Let $R$ be a random variable denoting the relevance label of a query-document pair and $R$ can assume values from a discrete finite set $\phi$ = \{0,1,2,3...,r\}.  Also let the current query be $q$ and the total number of documents that needs to be ranked for the current query $q$ is $n$, let us denote this set by $D^{}_{q}$. 
To derive the formula of $E[DCG@k]$, we start with the definition of expectation in probability theory.

\begin{equation*}
E[\rm DCG@k]\ = E\left[  \ \sum_{i=1}^{k}\frac{2^{R_i}\ -\ 1}{\log_b(i+1)} \right] = \sum_{i=1}^{k}\frac{E\left[2^{R_i}\ -\ 1\right]}{\log_b(i+1)}
\vspace{1mm}
\end{equation*}

So, the computation of $E[DCG@k]$ is based on the computation of $E[{2^{R_i}} - 1]$, which is the expected relevance label of the retrieved document at position $i$. Below we show how to estimate $E[{2^{R_i}} - 1]$ and first begin with the definition of expectation.

\begin{equation*}
E[2^{R_i}\ -1] = \ \sum_{j=0}^{r}(2^j - 1) \cdot Pr(R_i = j) 
\end{equation*}

Here, $Pr(R_i = j)$ is the probability that the retrieved document at position $i$ in a randomized ranking would assume a relevance label of $j$ with respect to the current query. Let us assume that $n_j$ be the number of documents with relevance label $j$, where $j \in \phi$, with respect to the current query. Thus, the constraint $\sum_{j=1}^r n_j = n$ holds, where $n$ is the total number of documents in $D_q$. Thus, $Pr(R_i = j)$ can essentially be computed by counting all the possible rankings which contain a document with relevance label $j$ (with respect to the current query) at position i and dividing it by the total number possible rankings up-to position $k$. Below we show the exact formula which is based on the permutation theory.

\begin{align*}
\boldsymbol{E[2^{R_i}\ -1]} &= \sum_{j=0}^{r}(2^j - 1) \cdot \left[\frac{^{n_j}_{}{P_1} \cdot ^{n-1}P_{k-1}  }{^{n}{P_k}}\right]  = \sum_{j=0}^{r}(2^j - 1) \cdot \left[\frac{ \frac{n_j !} {(n_j - 1)!} \cdot  \frac{(n-1)!}{(n-k)!}}    {\frac{n!}{(n-k)!}}\right]\\
&= \sum_{j=0}^{r}(2^j - 1)\cdot \left(\frac{n_j}{n}\right)= \sum_{j=0}^{r}(2^j - 1)\cdot Pr(R=j)= \boldsymbol{E[{2^R} - 1]}
\end{align*}

Note that, $E[2^R - 1]$ is different from $E[2^{R_i} - 1]$ because the former is independent of the position of a document in the ranked list, while later is dependent. However, the above derivation reveals that $E[2^{R_i} - 1]$ is indeed independent of the position $i$ and equals to $E[2^R - 1]$ for any $i$. Thus, the final formula for computing $E[DCG@k]$ boils down to the following formula:

\begin{equation}\label{equ:expected_nDCG}
    E[DCG@k] = E[2^{R} - 1] \cdot \sum_{i = 1}^{k}\frac{1}{log_2(i+1)}
\end{equation}

\begin{table}[!htb]\small
	\begin{center}
	    %\begin{adjustbox}{width=\linewidth}
		\begin{tabular}{c||c|c|c|c|c}\hline
		& \multicolumn{5}{c} {nDCG@}\\\cline{2-6}
		{\bf Method} & {\bf 5} & {\bf 10} & {\bf 15} & {\bf 20} & {\bf 30}\\\hline\hline
		ARank & 0.3218  & 0.3492  & 0.3704  & 0.3896 & 0.4237 \\\hline
		LNet & 0.1534  & 0.1827  & 0.2066  & 0.2288  & 0.2686\\\hline
		RBoost & 0.3062   & 0.3346  & 0.3578  & 0.3777  & 0.4141 \\\hline
		RF & 0.3832  & 0.4118  & 0.4325  & 0.4493  & 0.4795 \\\hline
		RNet & 0.154  & 0.1833  & 0.207  & 0.2292  & 0.269 \\\hline
		CA & 0.3985  & 0.4138  & 0.4288 & 0.4428  & 0.4707 \\\hline
		L2LR & 0.1977  & 0.2371  & 0.2696  & 0.2974  & 0.3444\\\hline
		LMART &  0.4365 &  0.454  & 0.4706 &  0.4856  &  0.513\\\hline
		\end{tabular}
		%\end{adjustbox}
		\caption{$nDCG$ scores of different LETOR methods for variable $k$ on MSLR-WEB30K data-set.}
% 		(Dataset and method details can be found in section~\ref{sec:CaseStudy})
% 		 Dataset MSLR-WEB30K
		\label{table:originalnDCG_MSLR}
		\end{center}
        \vspace{-1mm}
	    \begin{center}
		\begin{tabular}{c||c|c|c|c|c}\hline
		& \multicolumn{5}{c} {nDCG@}\\\cline{2-6}
		{\bf Method} & {\bf 5} & {\bf 10} & {\bf 15} & {\bf 20} & {\bf 30}\\\hline\hline
		ARank & 0.3881  & 0.4156  & 0.448  & 0.4797 & 0.5372 \\\hline
		LNet & 0.3767  & 0.4035  & 0.4384  & 0.4687  & 0.5282\\\hline
		RBoost & 0.3834   & 0.414  & 0.449  & 0.4807  & 0.5355 \\\hline
		RF & 0.4035  & 0.4286  & 0.4609  & 0.4914  & 0.5476 \\\hline
		RNet & 0.3809  & 0.4131  & 0.4451  & 0.4764  & 0.536 \\\hline
		CA & 0.3928  & 0.4207  & 0.4544 & 0.4824  & 0.5399 \\\hline
		L2LR & 0.3873  & 0.4159  & 0.4474  & 0.4779  & 0.538\\\hline
		LMART &  0.3931 &  0.4206  & 0.4535 &  0.4857  &  0.5441\\\hline
		\end{tabular}
		%\end{adjustbox}
		\caption{$nDCG$ scores of different LETOR methods for variable $k$ on MQ2007 data-set.}
% 		 Dataset MSLR-WEB10K
		\label{table:originalnDCG_MQ2007}
	\end{center}
	\vspace{-6mm}
\end{table}

\begin{table*}\footnotesize
	\begin{adjustbox}{width=0.9\linewidth}
		\begin{tabular}{c|c|c|c|c|c||c|c|c|c|c}\hline
		& \multicolumn{5}{c||} {$DCG^{U L}_{V_1}@$} & \multicolumn {5}{c} {$DCG^{U L}_{V_2}@$} \\\cline{2-11}
		{\bf Method} & {\bf 5} & {\bf 10} & {\bf 15} & {\bf 20} & {\bf 30} & {\bf 5} & {\bf 10} & {\bf 15} & {\bf 20} & {\bf 30}\\\hline\hline
		ARank & 0.249 & 0.2616  & 0.2719  & 0.2818 & 0.2995 & 0.2374  & 0.2531 & 0.2648 & 0.2761 & 0.2964 \\\hline
		
		LNet & 0.0977 & 0.1124  & 0.1257  & 0.1384  & 0.1617 & 0.0469 & 0.0606  & 0.0721  & 0.0837  & 0.1048   \\\hline
		
		RBoost & 0.2327   & 0.2474 & 0.2601  & 0.2708  & 0.2907 & 0.2217   & 0.2374  & 0.2509  & 0.2626  & 0.2852\\\hline
		
		RF & 0.3043  & 0.3187  & 0.3289 & 0.3365  & 0.3505 & 0.3086  & 0.3265  & 0.3385  & 0.3478 & 0.3652\\\hline
		
		RNet & 0.0982 & 0.113  & 0.126  & 0.1388  & 0.1621 & 0.0476  & 0.0614 & 0.0727  & 0.0843  & 0.1054   \\\hline
		
		CA & 0.3188  & 0.3208  & 0.3255 & 0.3306  & 0.3424 & 0.3259  & 0.3286  & 0.3341  & 0.3401  & 0.3545\\\hline
		
		L2LR & 0.1373  & 0.1605 & 0.1809  & 0.1985  & 0.2278 & 0.0982  & 0.1248  & 0.1475  & 0.1671  & 0.1996  \\\hline
		
		LMART & 0.3549  &  0.3588  &  0.3648 &  0.3706  & 0.3818 & 0.3677 &  0.3742 &  0.3824  &  0.3902 & 0.4056  \\\hline
		\end{tabular}
	\end{adjustbox}
		\caption{Upper \& Lower Bound Normalized DCG ($V_1$,$V_2$) scores of different LETOR methods for variable $k$: Each cell shows a particular $DCG^{U L}_{V}$ score with a particular $k$ on MSLR-WEB30K data-set.}
		\label{table:UL-nDCG_averageVaryKMSLR}
\end{table*}

\begin{table*}
\begin{adjustbox}{width=0.9\textwidth}
		\begin{tabular}{c|c|c|c|c|c||c|c|c|c|c}\hline
		& \multicolumn{5}{c||} {$DCG^{U L}_{V_1}@$} & \multicolumn {5}{c} {$DCG^{U L}_{V_2}@$}  \\\cline{2-11}
		{\bf Method} & {\bf 5} & {\bf 10} & {\bf 15} & {\bf 20} & {\bf 30} & {\bf 5} & {\bf 10} & {\bf 15} & {\bf 20} & {\bf 30}\\\hline\hline
		ARank & 0.2882 & 0.2991  & 0.3157  & 0.3314 & 0.3558 & 0.1348  & 0.2092 & 0.2587 & 0.2995 & 0.3638  \\\hline
		
		LNet & 0.2777 & 0.2886  &  0.3068 & 0.3213  &0.3481  &0.1141   & 0.1872  & 0.2485  & 0.284  & 0.3453  \\\hline
		RBoost & 0.2822 & 0.2975 & 0.3161  & 0.3317  &0.3542  & 0.1359   & 0.2061  & 0.2633  & 0.3042  &0.3635  \\\hline
		
		RF & 0.2992  & 0.3095 & 0.3262  & 0.3409 & 0.3642  & 0.1681  & 0.2356  & 0.2859  & 0.3223 & 0.3866 \\\hline
		
		RNet & 0.2791 & 0.2957 & 0.3133  & 0.3271  & 0.3536 & 0.1177  & 0.2044 & 0.2554  & 0.2908  & 0.3573  \\\hline
		
		CA & 0.2911  & 0.3031  & 0.32 & 0.3335  & 0.3582 & 0.1512  & 0.2214  & 0.2762  & 0.3029  & 0.3666  \\\hline
		
		L2LR & 0.2858  & 0.2991 & 0.315  & 0.3295  & 0.3562 & 0.1331  & 0.2097  & 0.2599  & 0.3001  & 0.368 \\\hline
		
		LMART & 0.2901  &  0.3019  &  0.3192 &  0.3355  & 0.361  & 0.1636 &  0.2311 &  0.2806  &  0.3183 & 0.3801 \\\hline
		\end{tabular}
	\end{adjustbox}
		\caption{Upper \& Lower Bound Normalized DCG ($V_1$,$V_2$,) scores of different LETOR methods for variable $k$: Each cell shows a particular $DCG^{U L}_{V}$ score with a particular $k$ on MQ2007 data-set.}
		\label{table:UL-nDCG_averageVaryKMQ2007}
		\vspace{-3mm}
\end{table*}

\subsection{Case-Study Observations}\label{ndcg_case_study}
This section discusses some observed differences between the original $nDCG$ and proposed $DCG^{U L}$.
For deeper analysis, we also created two special sub-sets of queries, i.e., 1) \textit{Uninformative} query-set and 2) \textit{Ideal} query-set, based on how close their average (of eight LETOR method and five cut-off k) Expected \textit{nDCG} is to the average real \textit{nDCG}. To achieve this, we computed both average Expected \textit{nDCG} and average real \textit{nDCG} for eight LETOR method and five different cut-offs. Specifically, we followed the steps from  Karmaker et.al.~\cite{DBLP:conf/cikm/KarmakerSZ20} to compute average real \textit{nDCG}. Table~\ref{table:originalnDCG_MSLR} and~\ref{table:originalnDCG_MQ2007} summarize the average (original) $nDCG$ scores of different LETOR methods for different values of $k$, i.e., $k = [5,10,15,20,30]$ for ``MSLR-WEB30K'' and ``MQ2007'' data-sets, respectively. One general observation from Table~\ref{table:originalnDCG_MSLR} and~\ref{table:originalnDCG_MQ2007} is that average $nDCG@k$ obtained by each method increases as we increase $k$ and the extent of this change is indeed significant. For example, RankNet achieves $nDCG$ value of $0.154$ and $0.269$ for $k = 5$ and $k = 30$ respectively with an increase of $74.6\%$ (Table~\ref{table:originalnDCG_MSLR}, ``MSLR-WEB30K'' data-set). 

\begin{figure}[!htb]
    \centering
    \includegraphics[width=0.8\linewidth]{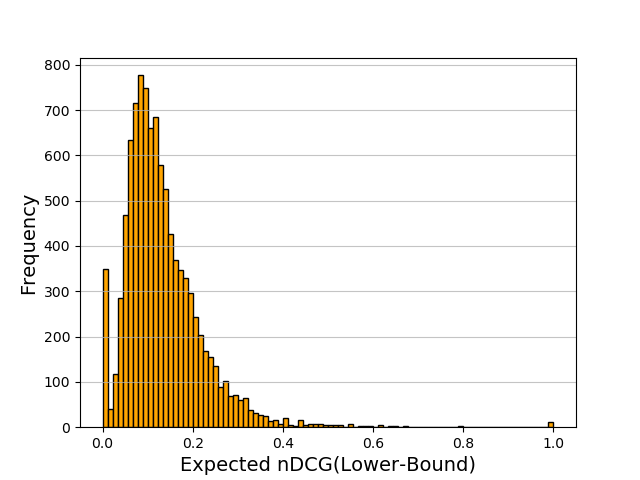}
    \vspace{-2mm}
    \caption{Histogram of expected $nDCG$ scores of $10,000$ queries from the ``MSLR-WEB30K'' data-set}
    \label{fig:expected_histogram}
    \vspace{-2mm}
\end{figure}

Next, we computed the expected $nDCG$ score for each query according to equation~\ref{equ:expected_nDCG}. Figure~\ref{fig:expected_histogram} shows the histogram of expected $nDCG$ scores of $10,000$ queries from the ``MSLR-WEB30K'' data-set. It is interesting to note that, a 
large portion of ``MSLR-WEB30K'' queries indeed demonstrates a large variance with high values in the ranges $[0.5 - 0.6]$. This justifies our position that lower-bound for each query can be very different and therefore, LB normalization should not be ignored while evaluating ranking performances.

Subsequently, we created two special sub-sets of queries based on the difference between their Expected \textit{nDCG} and the average real \textit{nDCG} obtained by eight LETOR methods, as defined below: 

\begin{itemize}[leftmargin=*,itemsep=0ex,partopsep=0ex,parsep=0ex]
    \item {\textbf{Uninformative Query-set:}} These are the top $1,000$ queries among the $10,000$ ``MSLR-WEB30K'' pool ($500$ in case of MQ-2007 data-set), where difference between the Expected \textit{nDCG} and the average real \textit{nDCG} is \textit{minimal}. In other words, these are the top $1,000$ ($500$) queries where the LETOR methods struggle to perform better than the random baseline.

    \item {\textbf{Ideal Query-set:}} These are the top $1,000$ queries among the $10,000$ ``MSLR-WEB30K'' pool ($500$ in case of MQ-2007 data-set), where difference between the Expected \textit{nDCG} and the average real \textit{nDCG} is \textit{maximal}. In other words, these are the top $1,000$ ($500$) queries where the LETOR methods outperforms the random baseline by the largest margin.
\end{itemize}

\subsubsection{\bf LB-normalized \textit{nDCG} yields different rankings compare to Original \textit{nDCG} for \textit{Uninformative} query-set:}
We first test whether our proposed metrics generate different ranking results compared with the original \textit{nDCG} or not.
Table~\ref{table:kendall_conflict_nDCG} shows the Kendall's $\uptau$ rank correlations between two rankings  induced by $nDCG$ and $DCG^{UL}$ scores in \textit{All}, \textit{Uninformative} or \textit{Ideal} query collections from the two data-sets. 
% Noted that \textit{Uninformative} and \textit{Ideal} are defined in section~\ref{sec:motivation}. 
We can notice that for both data-sets, $DCG^{U L}_{V_2}$ and $nDCG$ generate different rankings for Uninformative queries resulting the Kendall's $\uptau$ less than $1$ (i.e. $0.85$ and $0.928$). While for $DCG^{U L}_{V_1}$, it generate different rankings for Uninformative queries in `MSLR-WEB30K'' but not in ``MQ2007''.
Also, as expected in case of Ideal collections, there was no difference between $nDCG$ and $DCG^{UL}$ in both data-sets(Kendall's $\uptau$ is $1$).
Another interesting observation is while we use all query collection, only $DCG^{U L}_{V_2}$ generate different ranking result in case of ``MQ2007''.

\begin{table}[!htb]
	\begin{center}
	   
		\begin{tabular}{c||c|c|c|c}\hline
		& \multicolumn{4}{c} {\textbf{Kendall's $\uptau$}}\\\cline{2-5}
		{\bf Data-set} & \textbf{Version} &{\bf All }& {\bf uninform.} & {\bf Ideal} \\\hline\hline
		
		\multirow{2}{*}{\bf MSLR-WEB30K} &\multirow{2}{*}\textbf{nDCG vs V1} & 1 &\textbf{ 0.928} &    1 \\\cline{2-5}
		
		& nDCG vs V2 & 1 & \textbf{0.85} & 1 \\\hline\hline

		\multirow{2}{*}{\bf MQ2007} &\multirow{2}{*}\textbf{nDCG vs V1} & 1 & 1 &    1 \\\cline{2-5}
		
		& nDCG vs V2 & \textbf{0.785} & \textbf{0.928} & 1 \\\hline
	
		\end{tabular}

		\caption{Kendall's $\uptau$ rank correlations between LETOR method ranks based on $nDCG$ and two $DCG^{UL}$ on \textit{All}, \textit{uninformative} or \textit{ideal} query sets from two data-sets. }
		\label{table:kendall_conflict_nDCG}
		\end{center}
		\vspace{-3mm}
\end{table}

\subsubsection{\bf Statistical Significance Test Yields Different Outcomes for Original \textit{nDCG} Vs LB-normalized \textit{nDCG}:}Next we conducted statistical significance tests for every pair of LETOR methods based on their original $nDCG$ and $DCG^{U L}$ scores to see how many times the two metrics disagree on the relative performance between two competing LETOR methods. Specifically, we followed the \textit{bootstrap} Studentised Test (student's t-test) from Sakai~\cite{sakai2006evaluating} 
to verify whether the observed difference has occurred due to mere random fluctuations or not for each pair of LETOR methods. Using the most widely used confidence value of $0.05$ as the threshold, a p-value larger than $0.05$ means the two distributions are statistically same, otherwise the pair of distributions are statistically different. 
More specifically, we compared each pair of LETOR methods ($^8C_2=28$ pairs in total) with respect to five cut-off $k$, i.e., $k = [5,10,15,20,30]$. Thus, the total number of comparison is $28\times 5=140$.

Table~\ref{table:statistical_test_conflict_query_nDCG}  summarizes the number of disagreements between $nDCG$ and $DCG^{U L}$ in two data-sets. For instance, based on student's t-test, $DCG^{U L}_{V_2}$ disagreed with original $nDCG$ on $46$ (32\%) pairs of LETOR methods for \textit{Uninformative} query set from ``MSLR-WEB30K'', while \textbf{zero} disagreements for \textit{Ideal} query set. In ``MQ2007'', we can also observe $24$(17\%) pairs of disagreements for \textit{Uninformative} query set as well as there are 8 pairs of conflicts in \textit{Ideal} query set. In particularly, we also see $DCG^{U L}_{V_2}$ disagreed with original $nDCG$ on $6$ pairs for all query set from ``MQ2007''.

% which is consistent with Kendall's $\uptau$ results. 

Given the difference in outcomes and disagreements between the original $nDCG$ metric and it's LB-normalized version, a natural follow-up question now is: which metric is better? To answer this question, we compared the $nDCG$ and $DCG^{U L}$ metrics in terms of their \textit{Discriminative power} and \textit{Consistency}~\cite{sakai2006evaluating}. These are two popular methods for comparing evaluation measures.

\begin{table}[!htb]
	\begin{center}
	   
		\begin{tabular}{c||c|c|c|c}\hline
		& \multicolumn{4}{c} {\textbf{Conflict Cases}}\\\cline{2-5}
		{\bf Data-set} & \textbf{Version} &{\bf All }& {\bf uninform.} & {\bf Ideal} \\\hline\hline
		
		\multirow{2}{*}{\bf MSLR-WEB30K} &\multirow{2}{*}\textbf{nDCG vs V1} &  0 & \textbf{18 }&  0 \\\cline{2-5}
		
		& nDCG vs V2 & 0 & \textbf{46} & 0 \\\hline\hline

		\multirow{2}{*}{\bf MQ2007} &\multirow{2}{*}\textbf{nDCG vs V1} &0  & \textbf{20} & 1    \\\cline{2-5}
		
		& nDCG vs V2 & \textbf{6} & \textbf{24} & 8 \\\hline
	
		\end{tabular}

        \caption{We used Student's t-test to verify whether statistically significant difference occurred between a pair of LETOR methods while using $nDCG$ and $DCG^{U L}$ and counted the total number of disagreements on \textit{All}, \textit{uninformative} or \textit{ideal} query sets from two data-sets.}
		\label{table:statistical_test_conflict_query_nDCG}
		\end{center}
		\vspace{-3mm}
\end{table}

% \begin{table}[!htb]
% 	\begin{center}
	   
% 		\begin{tabular}{c||c|c|c}\hline
% 		& \multicolumn{3}{c} {\textbf{Conflict Cases}}\\\cline{2-4}
% 		{\bf Data-set} & {\bf version }& {\bf Closest} & {\bf Furthest} \\\hline\hline
		
% 		\multirow{2}{*}\textbf{MSLR-WEB30K} & nDCG and  V1 & 18  & 0    \\\cline{2-4}
		
% 		& nDCG and V2 &46 & 0 \\\hline\hline

% 		\multirow{2}{*} \textbf{MQ2007} & nDCG and V1 & 20 &  1 \\\cline{2-4}
% 		 & nDCG and V2 & 24 & 8 \\\hline
	
% 		\end{tabular}

% 		\caption{We used Student's t-test to verify whether statistically significant difference occurred between a pair of LETOR methods while using $nDCG$ and $DCG^{U L}$ and counted the total number of disagreements on ``MSLR-WEB30K'' and ``MQ2007'' data-set}
% 		\label{table:statistical_test_conflict_all_query}
% 		\end{center}
% 		\vspace{-3mm}
% \end{table}

% \label{sec:case4nDCG_ERR_MAP}

\subsubsection{\bf Distinguishability}\label{sec:nDCG_distin}~\\
We first focus on the implication of LB Normalization in terms of its capability to distinguish among multiple competing LETOR method pairs. To quantify distinguishability, we first utilize the \textit{discriminative} \textit{power}, which is a popular method for comparing evaluation metrics by performing a statistical significance test between each pair of LETOR methods and counting the number of times the test yields a significant difference. Note that \textit{discriminative power} is not about whether the metrics are right or wrong: it is about how often differences between methods can be detected with high confidence~\cite{sakai2011using}. We again follow Sakai~\cite{sakai2006evaluating} to use student's t-test to conduct this experiment and again use $0.05$ as our threshold. Using the aforementioned \textit{Uninformative} and \textit{Ideal} query collections, Table~\ref{table:student_discriminative_MSLR_WEB30K_nDCG} shows the total number of statistically significant differences that can be detected between pairs of LETOR methods in case of All queries, Uninformative queries and Ideal queries (from both data-sets), individually by the $nDCG$ and two $DCG^{UL}$ metrics.

\begin{table}[!htb]
	\begin{center}
	   
		\begin{tabular}{c||c|c|c|c}\hline
		& \multicolumn{4}{c} {\textbf{Number of Stat-Sig difference}}\\\cline{2-5}
		{\bf Data-set} & {\bf Version }& {\bf All}  & {\bf uniform.} & {\bf Ideal} \\\hline\hline
		
		\multirow{3}{*}{\bf MSLR-WEB30K} & nDCG &133 &33  &   130  \\\cline{2-5}
		
		& V1 &133 & \textbf{51} & 130 \\\cline{2-5}
		& V2 &133 &\textbf{78} & 130 \\\hline\hline

		\multirow{3}{*} {\bf MQ2007} & nDCG & 0& 9 & 7  \\\cline{2-5}
		
		& V1 & 0 &\textbf{29} &  \textbf{8} \\\cline{2-5}
		
		 & V2 &\textbf{6}  & \textbf{33} & \textbf{15}\\\hline
	
		\end{tabular}

		\caption{Student T-test induced total number of statistically significant differences detected based on $nDCG$ and $DCG^{UL}$ on \textit{All}, \textit{uninformative} or \textit{ideal} query sets from two data-sets. }
        \label{table:student_discriminative_MSLR_WEB30K_nDCG}
		\end{center}
		\vspace{-3mm}
\end{table}

On \textit{``MSLR-WEB30K'' Uninformative} query set, $nDCG$ could detect only $33$ (23\%) significantly different pairs. In contrast, both two proposed $DCG^{U L}_{V_1}$ and $DCG^{U L}_{V_2}$ can detect more cases of significant differences. Additionally, $DCG^{U L}_{V_2}$ achieve the best performance which detected \textbf{78} (55\%) significantly different pairs on the same set. On the other hand, on  \textit{``MSLR-WEB30K''  Ideal} query-set, both $nDCG$ and two $DCG^{UL}$ detected  \textbf{130} significantly different pairs.
It is evident that, both two $DCG^{UL}$ can better distinguish between two LETOR methods than $nDCG$ on ``MSLR-WEB30K'' data-set, while not compromising distinguishability in case of \textit{Ideal} queries, which is desired. We also observed similar improvements by $DCG^{UL}$ in case of ``MQ2007'' data-set. More importantly, $DCG^{UL}$ not only improve the distinguishability in case of \textit{uninformative} query set, it can also detect more different cases while using \textit{All} query set (for $DCG^{U L}_{V_2}$) and \textit{Ideal} query set (for both $DCG^{UL}$), which is a bonus.

% however, with improvement in distinguishability for both  \textit{Ideal} and \textit{Uninformative} queries, which is a bonus. 

We also computed another metric to quantify distinguishability: \textit{Percentage Absolute Differences (PAD)}. More specifically, we computed the percentage absolute differences between pairs of LETOR methods in terms of their original \textit{nDCG} and $DCG^{U L}$ scores, separately. The intuition here is that metrics with higher distinguishability will result in higher percentage absolute differences between pairs of LETOR methods. To elaborate, we first calculated the average value of both $nDCG$ and $DCG^{UL}$ with varying $k$ ( $k=\{5,10,15,20,30\}$ ) for each LETOR method, and then, computed the percentage absolute difference between each pair of LETOR methods in terms of those two metrics separately (one percentage for $nDCG$ and another for $DCG^{U L}$), then we calculated the average of those percentage absolute differences. This experiment was performed on both data-sets. Mathematically, we used the following formula for percentage absolute differences (PAD) in terms of original $nDCG$:

\vspace{-2mm}
{
\begin{equation}\label{equ:pad}
    PAD(nDCG)=\frac{|nDCG^{avg}_{M_1}-nDCG^{avg}_{M_2}|}{ \max\left(nDCG^{avg}_{M_1},nDCG^{avg}_{M_2}\right)} \times 100\%
\end{equation}}

Here, $M_1$ and $M_2$ are two different LETOR methods and $nDCG^{avg}_{M_1}$ is the average $nDCG$ score obtained by method $M_1$ with respect to varying $k$. The equation for $PAD(DCG^{U L})$ is similar thus omitted. Besides, we use this equation for the PAD calculation of our second case-study. Table~\ref{tab:PAD_result} shows these average percentage absolute differences of all possible LETOR method pairs in terms of original $nDCG$ and $DCG^{U L}$ scores on our two data-sets. 

% \textit{Note that, because we doubled the range of $DCG^{U L}$ from [0,1] to [-1, 1], in order to make a fair comparison, we divided the PAD score of $DCG^{U L}$ and presented them in Table~\ref{tab:PAD_result}}.

From this table, we can observe that while using $DCG^{U L}$, the PAD score of $DCG^{U L}$ is higher than the same for original $nDCG$ for all types of query collections, i.e., using \textit{All} queries, \textit{Uninformative} and \textit{Ideal} query sub-sets. For instance, the average PAD of $nDCG$ on ``MQ2007'' is \textbf{1.74}; while for  $DCG^{U L}_{V_2}$, the score is \textbf{6.42} (using all query). Similarly, we discovered that for \textit{Uninformative} query-set, $DCG^{U L}$ achieves a significant boost compared to the same in \textit{Ideal} query-set in both data-sets.
% The percentage increase obtained by $DCG^{U L}$ is around and  \textbf{26\%} w.r.t. original $nDCG$. On our second data-set ``MQ2007'', although the average PAD numbers are smaller than their corresponding values on ``MSLR-WEB10K'' (average PAD of  $DCG^{U L}$ is \textbf{3.21}), the increasing percentages are  \textbf{84\%}, which is higher than the results on ``MSLR-WEB10K''. 

These results show that the proposed LB normalization enhances the distinguishability of the original nDCG metric and can differentiate between two competing LETOR methods with a larger margin, which is a nice property of LB normalization.

\begin{table*}[!htb]\small
    \begin{adjustbox}{width=0.8\linewidth}
    \centering
    \begin{tabular}{c|c|c||c|c||c|c}\hline 

    & \multicolumn{6}{c}{ \bf PAD score}
    \\\cline{2-7}
    & \multicolumn{2}{c||} {\bf All Query} & \multicolumn {2}{c||} {\bf uninform} & \multicolumn {2}{c} {\bf Ideal} \\\cline{2-7}
    {\textbf{Metrics}} & {\bf MSLR} & {\bf MQ2007} & {\bf MSLR} & {\bf MQ2007} & {\bf MSLR} & {\bf MQ2007}\\ \hline\hline
    
    {\bf  nDCG}  &  31 & 1.74 &  7.39 & 5.85 &  35.74 & 1.61\\[1ex] \hline
    
    {$\bf DCG^{UL}_{V_1}$} &  {\bf35.7 } & {\bf 3.6} &  {\bf 9.98 } & {\bf7.825 } &  {\bf 40.21} & {\bf 1.98} \\[1ex] \hline

    {$\bf DCG^{UL}_{V_2}$} &  {\bf 46.7} & {\bf 6.42} &  {\bf 41.75} & {\bf 44.81} &  {\bf 44.53} & {\bf 2.98} \\[1ex] \hline

    %  {\bf  \makecell{Percentage Change \\ between nDCG and $\bf DCG^{U L}$}}  & \textbf{ +84\%} & \textbf{+26\%}\\\hline

    \end{tabular}
    \end{adjustbox}
    \caption{Percentage Absolute Difference between pairs of LETOR methods in terms of average $nDCG$ and $DCG^{U L}$ scores on \textit{All}, \textit{uninformative} or \textit{ideal} query sets from two data-sets.}
    \label{tab:PAD_result}
    \vspace{-5mm}
\end{table*}

\subsubsection{\bf Consistency}~\\
This experiment focuses to compare the relative ranking of LETOR methods in terms of their $nDCG$ and $DCG^{UL}$ scores, separately, \textit{across} different data-sets (``MQ2007'' Vs ``MSLR-WEB30K'') as well as \textit{across} \textit{Uninformative} and \textit{Ideal} query collections within the same data-set. The goal here is to see which metric yields a more stable ranking of LETOR methods across various types of documents and queries as well as across diverse set of data-sets. We computed \textit{swap rate}~\cite{sakai2006evaluating} to quantify the consistency of rankings induced by $nDCG$ and $DCG^{UL}$ metrics across different data-sets. The essence of swap rate is to investigate the probability of the event that two experiments are contradictory given an overall performance difference.

Table~\ref{table:Swap_Rate_across_dataset_NDCG} shows our swap rate results for $nDCG$ and $DCG^{UL}$ across the two data-sets, ``MSLR-WEB30K'' and ``MQ2007''. Note that in our original setup, we selected Uninformative/ Ideal 1000 queries from ``MSLR-WEB30K''. To make our results comparable, in this experiment we select 500 Uninformative/Ideal query from ``MSLR-WEB30K'' and compare the ranking result with the one from ``MQ2007''.  It can be observed that, both $nDCG$ and $DCG^{UL}$ share an identical swap rate probability when we conduct the experiment on the All/Uninformative/Ideal query collection (swap rate across data-sets is $0.107$, $0.42$ and $0.35$ for both metrics).

\begin{table}[!htb]
	\begin{center}
	   
		\begin{tabular}{c||c|c|c}\hline
		& \multicolumn{3}{c} {\textbf{Swap Rate}  }\\\cline{2-4}
		{\bf Metric}  & \textbf{All} & {\bf Uninform.} & {\bf Ideal}\\\hline\hline
		{\bf nDCG} & 0.107 & 0.42  & 0.35  \\[1ex] \hline
		
		$\bf DCG^{UL}_{V_1}$ & 0.107 &   0.42 &0.35 \\[1ex] \hline
		$\bf DCG^{UL}_{V_2}$ & 0.107 &  0.42  & 0.35\\[1ex] \hline
	
		\end{tabular}

		\caption{Swap rates between method ranks on \textit{All}/ \textit{uniform}/\textit{Ideal} queries across ``MSLR-WEB30K'' and ``MQ2007''  data-sets.}
        \label{table:Swap_Rate_across_dataset_NDCG}
		\end{center}
		\vspace{-3mm}
\end{table}

Table~\ref{table:Swap_Rate_across_query_set} also shows our swap rate results for $nDCG$ and $DCG^{UL}$ across \textit{Uninformative} Vs \textit{Ideal} queries from the same data-set.  We can still observe that both $nDCG$ and $DCG^{UL}$ generate the identical swap rate probability when we compare the ranking results across \textit{Uninformative} and \textit{Ideal} sets, except for $DCG^{UL}_{V_1}$ (generate a higher swap rate in ``MSLR-WEB30K'').

% This two tables indicate that our proposed metric does not hurt the consistency of original $nDCG$.

\begin{table}[!htb]
	\begin{center}
	   
		\begin{tabular}{c||c|c}\hline
		& \multicolumn{2}{c} {\textbf{Swap Rate}  }\\\cline{2-3}
		{\bf Metric}  & {\bf MSLR-WEB30K} & {\bf MQ2007}\\\hline\hline
		{\bf nDCG} & 0.21  & 0.5  \\[1ex] \hline 
		$\bf DCG^{UL}_{V_1}$ &  0.25  & 0.5\\[1ex] \hline
		
		$\bf DCG^{UL}_{V_2}$ &  0.21  & 0.5\\[1ex] \hline
	
		\end{tabular}
		\caption{Swap rates between method ranks on \textit{MSLR-WEB30K}/\textit{MQ2007} data-sets across ``uninformative'' and ``Ideal'' query collections.}
        \label{table:Swap_Rate_across_query_set}
		\end{center}
		\vspace{-3mm}
\end{table}

%% file: sections/7MAP.tex
\subsection{Case Study 2: MAP with Joint Upper \& Lower Bound Normalization}

\begin{figure}[!htb]
    \centering
    \includegraphics[width=0.8\linewidth]{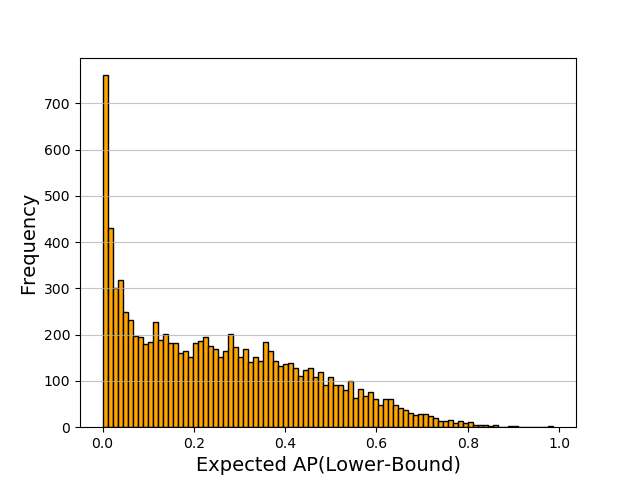}
    \vspace{-2mm}
    \caption{Histogram of expected $AP$ scores of $10,000$ queries from the ``MSLR-WEB30K'' data-set}
    \label{fig:expected_histogram_AP}
    \vspace{-2mm}
\end{figure}

For our second case study, we selected another popular evaluation metric called Mean Average Precision ($MAP$).
However, original $MAP$ computation needs binary label while our two data-sets are multi-relevance label. For consistency, in this paper, we only consider $0$ relevance score as negative and others are positive for both two data-sets.
Table ~\ref{table:MAPaverageVaryK_MSLR} and ~\ref{table:MAPaverageVaryK_MQ2007} show the original $MAP$ scores from two data-sets. 
Below, we will first present how we can compute a realistic lower bound for \textit{Sum Precision} ($SP$) by computing its expected value in case of a randomly ranked list of documents. Then, demonstrate our findings of lower bound normalized MAP. Again, lower bound normalized MAP essentially means upper lower bound normalized \textit{MSP}.

First, we also show the histogram of expected AP score for 10,000 queries from ``MSLR-WEB30K'' data-sets. Figure ~\ref{fig:expected_histogram_AP} shows the histogram of expected $AP$ scores of $10,000$ queries from the ``MSLR-WEB30K'' data-set. We can still observe that a large variance of high expected AP appeared in this data-set, indicating that can not be ignored.
Noted that we again created two special sub-sets of queries based on the difference between their Expected $AP$ and and average real $AP$ obtained by eight LETOR methods to define \textbf{Uninformative query-set} and \textbf{Ideal Query-set}( Details in ~\ref{ndcg_case_study}).

\subsubsection{\bf Lower Bound of SP (SP for Random Ranking):}
Given a query $q$, assume that $N_{p}$ is the total number of relevant documents , $N_{n}$ is the number of non-relevant document for query $q$. Also, assume $N_{p} > k$  and  $N_{n} > k$, $k$ is the cutoff variable. $Prec(i)$ is the precision at position $i$ and $R_{i}$ is the relevance at position $i$. Then, expectation of $SP@k$ in case of random ranking is the following:

$$E[SP@k] = \sum_{i=1}^{k} E[Prec(i) \cdot R_{i}]$$
\vspace{1mm}
Now assuming $Prec(i)$ and $R_{i}$ are independent, we have 
\vspace{1mm}
$$E[SP@k] = \sum_{i=1}^{k} E[Prec(i)] \cdot E[R_{i}]\mbox{, where,}$$ \vspace{1mm}
\[E[R_{i}] = P[R_{i} = 1]  \cdot 1 + P[R_{i} = 0] \cdot 0 = P[R_{r} = 1] = \frac{N_{p}}{N_{p} + N_{n}}\]\vspace{1mm}

\vspace{-2mm}
{
\begin{align*}
E[Prec@i] &= \frac{1}{i}\left[P\left(Prec@i= \frac{1}{i}\right)\right]
+ \frac{2}{i}\left[P\left(Prec@i = \frac{2}{i}\right)\right] + ... + \frac{i}{i}\left[P\left(Prec@i = \frac{i}{i}\right)\right] \\
&= \left(\frac{1}{i}\right) \left[\frac{\binom{N_{p}}{1}{\binom{N_{n}}{i-1}}}{\binom{N_{p} + N_{n}}{i}}\right]  +\left(\frac{2}{i}\right)\left[\frac{\binom{N_{p}}{2}{\binom{N_{n}}{i-2}}}{\binom{N_{p} + N_{n}}{i}}\right] + ... + \left(\frac{i}{i}\right)\left[\frac{\binom{N_{p}}{i}{\binom{N_{n}}{i-i}}}{\binom{N_{p}+N_{n}}{i}}\right]
=\left(\frac{1}{i}\right) \frac{1}{\binom{N_{p} + N_{n}}{i}} \sum_{j=1}^{i} j \binom{N_{p}}{j} \binom{N_{n}}{i-j}
\end{align*}}

% \begin{align*}
% E[Prec @i] = (\frac{1}{i}) \frac{1}{\binom{N_{p} + N_{n}}{i}} \sum_{j=1}^{i} j \binom{N_{p}}{k} \binom{N_{n}}{i-j}  \\
% \end{align*}

We will later prove that, 

\[\sum_{j=1}^{i} j \binom{N_{p}}{j} \binom{N_{n}}{i-j} = \frac{N_{p}}{N_{p} + N_{n}} i \binom{N_{p} + N_{n}}{i}\]

Thus, $E[Prec@i] = \frac{N_{p}}{N_{p} + N_{n}}$, Hence:

\vspace{-2mm}
{
\begin{align*}
 E[SP@k] & = \sum_{i=1}^{k} E[Prec(i)] \cdot E[R_{i}]= \sum_{i=1}^{k} \left(\frac{N_{p}}{N_{p} + N_{n}}\right)^2 = k \left(\frac{N_{p}}{N_{P} + N_{n}}\right)^{2}
\end{align*}}
\vspace{-2mm}

% \subsection*{\bf Proof of \textcircled{1}:}
Now, we will use induction to prove the following:

\vspace{-2mm}
{
\begin{equation} \label{equ: MAP2}
\sum_{j=1}^{i} j \binom{N_{p}}{j} \binom{N_{n}}{i-j} = \left(\frac{N_{p}}{N_{p} + N_{n}}\right) i \binom{N_{p} + N_{n}}{i} 
\end{equation}}

\vspace{-2mm}
{
\textbf{Base case:} For i = 1, L.H.S = $1 \binom{N_{p}}{1} \binom{N_{n}}{1-1} = N_{p}$}

{
\begin{equation*}
R.H.S= \left(\frac{N_{p}}{N_{p} + N_{n}}\right) 1 \binom{N_{p} + N_{n}}{1}= \frac{N_{p}}{N_{p} + N_{n}} \left(N_{p} + N_{n}\right) = N_{p} \\
\end{equation*}}

\vspace{-1mm}
So, equation \ref{equ: MAP2} is true for $i$ = 1

\textbf{Induction step:} Now, Let's assume equation \ref{equ: MAP2} is true for $i$ = $i$-1, then we get the following:

\begin{equation} \label{equ:MAP3}
\sum_{j=1}^{i-1} j \binom{N_{p}}{j} \binom{N_{n}}{i-1-j} = \frac{N_{p}}{N_{p} + N_{n}} (i-1) \binom{N_{p} + N_{n}}{i-1}
\end{equation}

% \resizebox{.9\hsize}

\begin{align*}
    % &\resizebox{.8\hsize}{!}{$A+B+C+D+E+F+G+H+I+J+K+L+M+N+O+P+Q+R+S+T+U+V+W+X+Y+Z$}\\
    &\mbox{Now, }\boldsymbol{\sum_{j=1}^{i} j \binom{N_{p}}{j} \binom{N_{n}}{i-j}}=\sum_{j=1}^{i-1} j \binom{N_{p}}{j} \binom{N_{n}}{i-j} + i\binom{N_{p}}{i}\\
    &= \sum_{j=1}^{i-1} j \binom{N_{p}}{j} [\binom{N_{n}+1}{i- j} - \binom{N_{n}}{i-j-1}] + i \binom{N_{p}}{i}\\
    &=  \left[\sum_{j=1}^{i-1} j \binom{N_{p}}{j} \binom{N_{n} + 1}{i-j}\right] + i \binom{N_{p}}{i}  - \left[\sum_{j=1}^{i-1} j \binom{N_{p}}{j} \binom{N_{n}}{i-j-1}\right]\\
    &= \sum_{j=1}^{i} j \binom{N_{p}}{j} \binom{N_{n} + 1}{ i - j} - \left(\frac{N_{p}}{N_{p} + N_{n}}\right) (i-1) \binom{N_{p} + N_{n}}{i-1} \hspace{10mm} \mbox{[From (\ref{equ:MAP3})]} \\
    % &= \sum_{j=1}^{i} j \binom{N_{p}}{j} \binom{N_{n} + 1}{i-j} - (\frac{N_{p}}{N_{p} + N_{n}}) (i-1)\binom{N_{p} + N_{n}}{i-1} \\
    &= \sum_{j=1}^{i} {N_{p}} \binom{N_{p} - 1}{j - 1} \binom{N_{n} + 1}{i-j} - \left(\frac{N_{p}}{N_{p} + N_{n}}\right) (i-1)\binom{N_{p} + N_{n}}{i-1} 
    \hspace{10mm} As, \left[{j\binom{N_{p}}{i} = N_{p} \binom{N_{p} - 1}{j-1}}\right] \\
    &= N_{p}\sum_{j=1}^{i}\binom{N_{p} -1}{j-1}\binom{N_{n} + 1}{i-j} - \left(\frac{N_{p}}{N_{p} + N_{n}}\right) (i-1)\binom{N_{p} + N_{n}}{i-1}\\
    &=  N_{p} \binom{N_{p} + N_{n}}{i-1} - \left(\frac{N_{p}}{N_{p} + N_{n}}\right) (i-1) \binom{N_{p} + N_{n}}{i-1} \\ 
    % &= N_{p} \binom{N_{p} + N_{n}}{i-1} - \frac{N_{p}}{N_{p} + N_{n}} (i-1) \binom{N_{p} + N_{n}}{i-1} \\ 
    &= \binom{N_{p} + N_{n}}{i-1} \left(\frac{N_{p}}{N_{P} + N_{n}}\right) [N_{p} + N_{n} - i+1]
    =\left[(N_{p} + N_{n} -i+1)\binom{N_{p} + N_{n}}{i-1}\right]\left(\frac{N_{p}}{N_{p} + N_{n}}\right)\\
    &= \boldsymbol{i\binom{N_{p} + N_{n}}{i}\left(\frac{N_{p}}{N_{p} + N_{n}}\right)}
\end{align*}

Proof completed because 

\[(n-r+1)\binom{n}{r-1} = r\binom{n}{r}\]

\vspace{-4mm}
\begin{table}[!htb]\small
	\begin{center}
		\begin{tabular}{c||c|c|c|c|c}\hline
		& \multicolumn{5}{c} {MAP@}\\\cline{2-6}
		{\bf Method} & {\bf 5} & {\bf 10} & {\bf 15} & {\bf 20} & {\bf 30}\\\hline\hline
		ARank & 0.5414  & 0.4948  & 0.4724 & 0.4598 & 0.4493 \\\hline
		LNet & 0.3203  & 0.2994 & 0.293  & 0.2911  & 0.2943 \\\hline
		RBoost & 0.5449   & 0.4967  & 0.475  & 0.4618  & 0.452 \\\hline
		RF & 0.6216  & 0.5717  & 0.5433  & 0.5244  & 0.5053 \\\hline
		RNet & 0.3212  & 0.3008  & 0.2939  & 0.2919  & 0.2956\\\hline
		CA & 0.6235  & 0.5631  & 0.53  & 0.5107  & 0.4903 \\\hline
		L2LR & 0.356 & 0.3353 & 0.333 & 0.3353 & 0.3457 \\\hline
		LMART &  0.6487  &  0.5928 &  0.5613  &  0.5414 & 0.5198 \\\hline

		\end{tabular}
		\caption{$MAP$ scores of different LETOR methods for variable $k$ on  'MSLR-WEB30K' dataset. }
		\label{table:MAPaverageVaryK_MSLR}
% 		\vspace{-3mm}

		\begin{tabular}{c||c|c|c|c|c}\hline
		& \multicolumn{5}{c} {MAP@}\\\cline{2-6}
		{\bf Method} & {\bf 5} & {\bf 10} & {\bf 15} & {\bf 20} & {\bf 30}\\\hline\hline
		ARank & 0.3066  & 0.2923  & 0.302  & 0.3173 & 0.3624 \\\hline
		LNet & 0.3379  & 0.3233  & 0.3328  & 0.3468  & 0.3905\\\hline
		RBoost & 0.3467   & 0.3366  & 0.3477  & 0.3636  & 0.4035 \\\hline
		RF & 0.3674  & 0.352  & 0.3585  & 0.3736  & 0.414 \\\hline
		RNet & 0.3281  & 0.3175  & 0.3275  & 0.3443  & 0.3878 \\\hline
		CA & 0.3597 & 0.3457  & 0.356 & 0.3716  & 0.4127 \\\hline
		L2LR & 0.3543  & 0.3386  & 0.3458  & 0.3607  & 0.404\\\hline
		LMART &  0.3582 &  0.3459  & 0.3539 &  0.3692  &  0.4101\\\hline
		\end{tabular}
		%\end{adjustbox}
		\caption{$MAP$ scores of different LETOR methods for variable $k$ on  'MQ2007' dataset.}

		\label{table:MAPaverageVaryK_MQ2007}
	\end{center}
\end{table}

\begin{table*}\footnotesize
	\begin{adjustbox}{width=0.9\linewidth}
		\begin{tabular}{c|c|c|c|c|c||c|c|c|c|c}\hline
		& \multicolumn{5}{c||} {$MSP^{U L}_{V_1}@$} & \multicolumn {5}{c} {$MSP^{U L}_{V_2}@$}    \\\cline{2-11}
		{\bf Method} & {\bf 5} & {\bf 10} & {\bf 15} & {\bf 20} & {\bf 30} & {\bf 5} & {\bf 10} & {\bf 15} & {\bf 20} & {\bf 30} \\\hline\hline

		ARank & 0.3856  & 0.3387  & 0.3156 & 0.3017 & 0.2868 & 0.3472  & 0.3055  & 0.2799 & 0.2617 & 0.2373  \\\hline

		LNet & 0.1978  & 0.1732 & 0.1641  & 0.1597  & 0.1572  & -0.0721  & -0.0573 & -0.0508  & -0.0452  & -0.0389 \\\hline

		RBoost & 0.3905   & 0.3422 & 0.3196  & 0.305  & 0.2905 & 0.3502  & 0.3019  & 0.2754  & 0.2567  & 0.2339  \\\hline

		RF & 0.4579  & 0.4079  & 0.3791  & 0.3591  & 0.3363 & 0.4783  & 0.427 & 0.39  & 0.3608  & 0.322  \\\hline

		RNet & 0.1988  & 0.1745  & 0.1651  & 0.1606  & 0.1585 & -0.0718  & -0.0551  & -0.0498  & -0.0452  & -0.038 \\\hline

		CA & 0.4594  & 0.4001  & 0.3673  & 0.3471  & 0.3231 & 0.4836  & 0.4127  & 0.3676  & 0.3385  & 0.2977\\\hline

		L2LR & 0.226 & 0.2018 & 0.1963 & 0.1951 & 0.1982 & -0.0223 & -0.004 & 0.0149 & 0.0312 & 0.055  \\\hline

		LMART &  0.482  &  0.4265&  0.3948  &  0.3741 & 0.3488 &  0.5259  &  0.4632 & 0.4211  &  0.3896 & 0.3466 \\\hline
		\end{tabular}
		\end{adjustbox}
		\caption{Upper \& Lower Bound Normalized MSP ($V_1$,$V_2$) scores of different LETOR methods for variable $k$: Each cell shows a particular $MSP^{U L}_{V}$ score with a particular $k$. MSLR-WEB30K dataset.}
		\label{table:ULMAP1averageVaryK_MSLR}
\end{table*}

\begin{table*}
    \begin{adjustbox}{width=0.9\textwidth}

		\begin{tabular}{c|c|c|c|c|c||c|c|c|c|c}\hline
		& \multicolumn{5}{c||} {$MSP^{U L}_{V_1}@$} & \multicolumn {5}{c} {$MSP^{U L}_{V_2}@$}     \\\cline{2-11}
		{\bf Method} & {\bf 5} & {\bf 10} & {\bf 15} & {\bf 20} & {\bf 30} & {\bf 5} & {\bf 10} & {\bf 15} & {\bf 20} & {\bf 30}\\\hline\hline
		
		ARank & 0.2366 & 0.219  & 0.2222  & 0.2287 & 0.2478 & 0.0392  & 0.0778 & 0.0116 & 0.14 & 0.1905 \\\hline

		LNet & 0.2676 & 0.2492  &  0.2519 & 0.257  &0.2744  &0.0909   & 0.1315  & 0.1638  & 0.1846  & 0.2257   \\\hline

		RBoost & 0.2738 & 0.2603 & 0.2647  & 0.2714  &0.2853  & 0.123   & 0.154  & 0.188   & 0.213  &0.2513 \\\hline

		RF & 0.2914  & 0.2732 & 0.2739  & 0.28 & 0.2941  & 0.1586  & 0.1904  & 0.206  & 0.226 & 0.2729\\\hline

		RNet & 0.259 & 0.2443 & 0.2476  & 0.2552  & 0.2724 & 0.085  & 0.1308 & 0.1567  & 0.1825  & 0.222   \\\hline
		
		CA & 0. 2863 & 0.2689  & 0.2728 & 0.2794  & 0.2941 & 0.1422  & 0.1741  & 0.198  & 0.2204  & 0.2584 \\\hline
		
		L2LR & 0.2806  & 0.2622 & 0.2633  & 0.2693  & 0.2861 & 0.1232  & 0.1548  & 0.1846  & 0.2093  & 0.2543 \\\hline
		
		LMART & 0.2829  &  0.2673  &  0.2691 &  0.2755  & 0.2905  & 0.1541 &  0.1949 &  0.2138  &  0.2369 & 0.2725 \\\hline
		\end{tabular}
	\end{adjustbox}
		\caption{Upper \& Lower Bound Normalized MSP ($V_1$,$V_2$) scores of different LETOR methods for variable $k$: Each cell shows a particular $MSP^{U L}_{V}$ score with a particular $k$. MQ2007 dataset.}
		\label{table:ULMAP1averageVaryK_MQ2007}
		\vspace{-1mm}
\end{table*}

\vspace{2mm}
\subsubsection{\bf LB-normalized \textit{MAP} yields different rankings compare to Original \textit{MAP} for \textit{Uninformative} query-set:} 
Table~\ref{table:MAP_kendall_conflict} shows the Kendall's $\uptau$ rank correlations between two rankings  induced by $MAP$ and $MSP^{UL}$ scores in \textit{All}, \textit{Uninformative} or \textit{Ideal} query collections for the two data-sets. 
Firstly, we can notice that for both data-sets, $MSP^{U L}_{V_1}$ and $MAP$ generate identical rankings for different query set which indicate that there is no difference between $MAP$ with $MSP^{U L}_{V_1}$ in terms of  Kendall's $\uptau$ rank test. While for $MSP^{U L}_{V_2}$, it generate different rankings for all kinds of query collections in both two data-sets. 
For instance, in ``MQ2007'', Kendall's $\uptau$ correlation between $MAP$ and $MSP^{U L}_{V_2}$ are \textbf{0.785},\textbf{ 0.624} and \textbf{1} for \textit{all}, \textit{uninformative} and \textit{ideal} query set, suggesting that $MSP^{U L}_{V_2}$ achieves different outcomes. In addition, the impact is more prominent in case of \textit{uninformative} compared with \textit{ideal}. 

\begin{table}[!htb]
	\begin{center}
	   
		\begin{tabular}{c||c|c|c|c}\hline
		& \multicolumn{4}{c} {\textbf{Kendall's $\uptau$}}\\\cline{2-5}
		{\bf Data-set} & \textbf{Version} &{\bf All }& {\bf uninform.} & {\bf Ideal} \\\hline\hline
		
		\multirow{2}{*}{\bf MSLR-WEB30K}
		&\multirow{2}{*}\textbf{MAP vs V1} & 1  & 1 & 1   \\\cline{2-5}
		
		& MAP vs V2 & \textbf{0.928} &\textbf{ 0.857} & \textbf{0.928} \\\hline\hline

		\multirow{2}{*}{\bf MQ2007} &\multirow{2}{*}\textbf{MAP vs V1} &1 & 1 &  1  \\\cline{2-5}
		
		& MAP vs V2 & \textbf{0.785} &\textbf{ 0.624} & 1 \\\hline
	
		\end{tabular}

		\caption{Kendall's $\uptau$ rank correlations between LETOR method ranks based on $MAP$ and two $MSP^{UL}$ on \textit{All}, \textit{uninformative} or \textit{ideal} query sets from two data-sets. }
		\label{table:MAP_kendall_conflict}
		\end{center}
		\vspace{-3mm}
\end{table}

\subsubsection{\bf Statistical Significance Test Yields Different Outcomes for Original \textit{MAP} Vs LB-normalized \textit{MAP}:} We again conducted statistical significance tests for every pair of LETOR methods based on their original $MAP$ and $MSP^{U L}$ scores to see how many times the two metrics disagree on the relative performance between two competing LETOR methods.

Table~\ref{table:statistical_test_conflict_MAP}  summarizes the number of disagreements between $MAP$ and $MSP^{U L}$ in two data-sets. For instance, based on student's t-test, $MSP^{U L}_{V_2}$ disagreed with original $MAP$ on $36$ (26\%) pairs of LETOR methods for \textit{Uninformative} query set from ``MSLR-WEB30K'', while \textbf{4} disagreements for \textit{Ideal} query set. Although none of $MSP^{U L}$ disagree with original $MAP$ while using \textit{All} query set from ``MSLR-WEB30K'',  there are still $1$ and $8$ conflicts appeared in ``MQ2007'' for two UL normalized version respectively.

% In ``MQ2007'', we can also observe $24$(17\%) pairs of disagreements for \textit{Uninformative} query set as well as there are 8 pairs of conflicts in \textit{Ideal} query set. In particularly, we also see $DCG^{U L}_{V_2}$ disagreed with original $nDCG$ on $6$ pairs for all query set from ``MQ2007'' which is consistent with Kendall's $\uptau$ results. 

Given the difference in outcomes and disagreements between the original $MAP$ metric and it's LB-normalized version, we still trying to compare these two metrics in terms of their \textit{Discriminative power} and \textit{Consistency} just like what we did in $nDCG$.

\begin{table}[!htb]
	\begin{center}
	   
		\begin{tabular}{c||c|c|c|c}\hline
		& \multicolumn{4}{c} {\textbf{Conflict Cases}}\\\cline{2-5}
		{\bf Data-set} & \textbf{Version} &{\bf All }& {\bf uninform.} & {\bf Ideal} \\\hline\hline
		
		\multirow{2}{*}{\bf MSLR-WEB30K} &\multirow{2}{*}\textbf{MAP vs V1} &  0 &\textbf{15}  &2   \\\cline{2-5}
		
		& MAP vs V2 & 0 & \textbf{36} & 4 \\\hline\hline

		\multirow{2}{*}{\bf MQ2007} &\multirow{2}{*}\textbf{MAP vs V1} &1  & \textbf{2} & 3    \\\cline{2-5}
		
		& MAP vs V2 &\textbf{ 8} & \textbf{21} & 17 \\\hline
	
		\end{tabular}

        \caption{We used Student's t-test to verify whether statistically significant difference occurred between a pair of LETOR methods while using $MAP$ and $MSP^{U L}$ and counted the total number of disagreements on \textit{All}, \textit{uninformative} or \textit{ideal} query sets from two data-sets. }
		\label{table:statistical_test_conflict_MAP}
		\end{center}
		\vspace{-3mm}
\end{table}

\subsubsection{\bf Distinguishability}~\\
We again follow Sakai~\cite{sakai2006evaluating} to use student's t-test to conduct this experiment and use $0.05$ as our threshold. Using the aforementioned \textit{Uninformative} and \textit{Ideal} query collections, Table~\ref{table:MAP_student_discriminative} shows some interesting results of these statistical tests for different query sets in `MSLR-WEB10K` and''``MQ2007'' data-sets.

\begin{table}[!htb]
	\begin{center}
	   
		\begin{tabular}{c||c|c|c|c}\hline
		& \multicolumn{4}{c} {\textbf{Number of Stat-Sig difference}}\\\cline{2-5}
		{\bf Data-set} & {\bf Version }& {\bf All}  & {\bf uniform.} & {\bf Ideal} \\\hline\hline
		
		\multirow{3}{*}{\bf MSLR-WEB30K} & MAP &  129 & 61 & 122    \\\cline{2-5}
		
		& V1 & 129 & \textbf{76} & 124 \\\cline{2-5}
		& V2 & 129  & \textbf{81 }& 122 \\\hline\hline

		\multirow{3}{*} {\bf MQ2007} & MAP &45 &0  & 71  \\\cline{2-5}
		
		& V1 &\textbf{50}  &\textbf{2} &  \textbf{74} \\\cline{2-5}
		
		 & V2 & \textbf{59} &\textbf{21}  & \textbf{88} \\\hline
	
		\end{tabular}

		\caption{Student T-test induced total number of statistically significant differences detected based on $MAP$ and $MSP^{UL}$ on \textit{All}, \textit{uninformative} or \textit{ideal} query sets from two data-sets. }
        \label{table:MAP_student_discriminative}
		\end{center}
		\vspace{-3mm}
\end{table}

On \textit{``MSLR-WEB30K'' Uninformative} query set, although $MAP$ detect  $61$ (43\%) significantly different pairs, both two proposed $MSP^{U L}_{V_1}$  and $DCG^{U L}_{V_2}$ can detect more cases of significant differences. What can be clearly seen is  $MSP^{U L}_{V_2}$ still achieve the best performance which detected \textbf{81} (57\%) significantly different pairs on the same set. On the other hand, on  \textit{``MSLR-WEB30K''  Ideal} query set, both $MAP$ and two $MSP^{UL}$ detected  around \textbf{122} significantly different pairs.
More interestingly, in ``MQ2007'', while original $MAP$ detect $45$ cases of different pairs using all query set, $MSP^{U L}$ indeed improve this performance (for $MSP^{U L}_{V_1}$ is $50$ and $MSP^{U L}_{V_2}$ is $59$).  Specifically in uninformative query set, $MAP$ can not detect any significantly different pairs. However, $MSP^{U L}_{V_2}$ can detect $21$ pairs of difference, which is very important. On the other hand, $MSP^{U L}_{V_2}$ can even detect more cases in $ideal$ query set.
It is evident that, both two $MSP^{UL}$ can better distinguish between two LETOR methods than $MAP$ on two data-sets, while not compromising distinguishability in case of \textit{Ideal} queries (even improve the distinguishability in ``MQ2007'').

Again, we use the formula ~\ref{equ:pad} to compute the percentage of absolute differences between pairs of LETOR methods in terms of their original $MAP$ and $MSP^{U L}_{V}$, separately. Here, {\bf X} represents $MAP$ and $MSP^{U L}_{V_{1,2}}$. (Details of PAD can be found in ~\ref{sec:nDCG_distin}).

\begin{table*}[!htb]
    \begin{adjustbox}{width=0.8\linewidth}
    \centering
    \begin{tabular}{c|c|c||c|c||c|c}\hline 

    & \multicolumn{6}{c}{ \bf PAD score}
    \\\cline{2-7}
    & \multicolumn{2}{c||} {\bf All Query} & \multicolumn {2}{c||} {\bf uninform} & \multicolumn {2}{c} {\bf Ideal} \\\cline{2-7}
    {\textbf{Metrics}} & {\bf MSLR} & {\bf MQ2007} & {\bf MSLR} & {\bf MQ2007} & {\bf MSLR} & {\bf MQ2007}\\ \hline\hline
    
    {\bf  MAP}  & 25.57  & 5.91 & 12.28  & 5.89 & 30.18  & 6.77\\\hline

    {$\bf MSP^{UL}_{V_1}$} & \textbf{31.84}  & \textbf{6.86} & \textbf{ 16} & \textbf{7.19} &\textbf{35.53}  & \textbf{8.04} \\[1ex]\hline

    {$\bf MSP^{UL}_{V_2}$} &  \textbf{97.63} & \textbf{20.01} & \textbf{25.65}  & \textbf{28.27} &  \textbf{48.29} & \textbf{13.49 }\\[1ex]\hline

    %  {\bf  \makecell{Percentage Change \\ between nDCG and $\bf DCG^{U L}$}}  & \textbf{ +84\%} & \textbf{+26\%}\\\hline

    \end{tabular}
    \end{adjustbox}
    \caption{Percentage Absolute Difference between pairs of LETOR methods in terms of average $MAP$ and $MSP^{U L}$ scores on \textit{All}, \textit{uninformative} or \textit{ideal} query sets from two data-sets..}
    \label{tab:MAP_PAD_result}
    \vspace{-5mm}
\end{table*}

Table~\ref{tab:MAP_PAD_result} illustrates the PAD score in case of $MAP$ and proposed two $MSP^{U L}$ from two data-sets for different query collections.

From this table, we can still observe that while using $MSP^{U L}$ can achieve higher PAD score than the same for original $MAP$ for all types of query collections, i.e., using \textit{All} queries, \textit{Uninformative} and \textit{Ideal} query sub-sets. For instance, the average PAD of $MAP$ on ``MSLR-WEB30K'' is \textbf{25.57}; while for  $MSP^{U L}_{V_2}$, the score is \textbf{97.63} (using all query). Similarly, we can still discovered that for \textit{Uninformative} query-set, both $MSP^{U L}$ versions achieve a significant boost compared to the same in \textit{Ideal} query set in both data-sets.

These results show that the proposed LB normalization 
again improve the distinguishability of original $MAP$ and can better differentiate between the quality of two LETOR methods  with a larger margin.

\subsubsection{\bf Consistency}~\\
This experiment again focuses to compare the relative ranking of LETOR methods in terms of their $MAP$ and $MSP^{UL}$ scores, separately, \textit{across} different data-sets (``MQ2007'' Vs ``MSLR-WEB30K'') as well as \textit{across} \textit{Uninformative} and \textit{Ideal} query collections within the same data-set. We computed \textit{swap rate}~\cite{sakai2006evaluating} to quantify the consistency of rankings induced by $MAP$ and $MSP^{UL}$ metrics across different data-sets.
Table~\ref{table:Swap_Rate_across_dataset_MAP} shows our swap rate results for $MAP$ and $MSP^{UL}$ across the two data-sets, ``MSLR-WEB30K'' and ``MQ2007''. 
In contrast to \textit{identical} swap rate scores in $nDCG$ and $DCG^{UL}$, $MSP^{UL}_{V_2}$ can achieve a overall  \textbf{lower} swap rate(swap rate of $MAP$ is $0.25$ while $0.178$ for $MSP^{UL}_{V_2}$) across a data-sets comparison while considering all query set.

% It can be observed that, both $nDCG$ and $DCG^{UL}$ share an identical swap rate probability when we conduct the experiment on the All/Uninformative/Ideal query collection (swap rate across data-sets is $0.107$, $0.42$ and $0.35$ for both metrics).

\begin{table}[!htb]
	\begin{center}
	   
		\begin{tabular}{c||c|c|c}\hline
		& \multicolumn{3}{c} {\textbf{Swap Rate}  }\\\cline{2-4}
		{\bf Metric}  & \textbf{All} & {\bf Uninform.} & {\bf Ideal}\\\hline\hline
		{\bf MAP} & 0.25 & 0.357  & 0.2857  \\\hline
		
		$\bf MSP^{UL}_{V_1}$ & 0.25 & 0.321   & 0.25\\[1ex]\hline
		$\bf MSP^{UL}_{V_2}$ & \textbf{0.178} &    \textbf{0.25}& 0.321\\[1ex]\hline
	
		\end{tabular}

		\caption{Swap rates between method ranks on \textit{All}/ \textit{uniform}/\textit{Ideal} queries across ``MSLR-WEB30K'' and ``MQ2007''  data-sets.}
        \label{table:Swap_Rate_across_dataset_MAP}
		\end{center}
		\vspace{-3mm}
\end{table}

Table~\ref{table:Swap_Rate_across_query_set_MAP} also shows our swap rate results for $MAP$ and $MSP^{UL}$ across \textit{Uninformative} Vs \textit{Ideal} queries from the same data-set. Similarly,  we can still observe that  $MSP^{UL}_{V_2}$ can obtain a more consistent ranking results across different query collection, which is very useful for an evaluation metric.

% both $nDCG$ and $DCG^{UL}$ generate the identical swap rate probability when we compare the ranking results across \textit{Uninformative} and \textit{Ideal} sets, except for $DCG^{UL}_{V_1}$ (generate a higher swap rate in ``MSLR-WEB30K'').

\begin{table}[!htb]
	\begin{center}
	   
		\begin{tabular}{c||c|c}\hline
		& \multicolumn{2}{c} {\textbf{Swap Rate}  }\\\cline{2-3}
		{\bf Metric}  & {\bf MSLR-WEB30K} & {\bf MQ2007}\\\hline\hline
		
		{\bf MAP} & 0.1428  & 0.3928  \\\hline
		$\bf MSP^{UL}_{V_1}$ &  0.1428  & 0.3928\\[1ex]\hline
		
		$\bf MSP^{UL}_{V_2}$ & \textbf{0.1071}    & \textbf{0.2857}\\[1ex]\hline
	
		\end{tabular}
		\caption{Swap rates between method ranks on \textit{MSLR-WEB30K}/\textit{MQ2007} data-sets across ``uninformative'' and ``Ideal'' query collections.}
        \label{table:Swap_Rate_across_query_set_MAP}
		\end{center}
		\vspace{-3mm}
\end{table}

%% file: sections/8Conclusions.tex
\section{Discussions and Conclusion}\label{sec:conclusion}

In this paper, we presented a novel perspective towards evaluation of Information Retrieval (IR) systems. Specifically, we performed two case-study on \textit{nDCG}, and \textit{MAP} both are widely popular metrics for IR evaluation, and started with the observation that, traditional \textit{nDCG} and \textit{MAP} computation does not include a query-specific lower-bound normalization although they include a query-specific upper-bound normalization. In other words, the current practice is to assume a uniform lower bound (zero) across all queries while computing \textit{nDCG} and \textit{MAP}, an assumption which is incorrect. This limitation raises a question mark on the previous comparative studies involving multiple ranking methods where an average evaluation metric score is reported, because \textit{Uninformative} vs. \textit{Informative} vs. \textit{Ideal} queries are rewarded equally in traditional IR evaluation metric computation and the expected lower-bound of the evaluation metric is ignored. \textit{How can we incorporate query-specific LB normalization into IR evaluation metrics and how will it impact IR evaluation in general?} This is the central issue we investigated in this paper.

\smallskip
\noindent\textbf{Conceptual Leap:} To address the aforementioned issue, we proposed to penalize the traditional IR evaluation metric score of each query with a lower-bound normalization term specific to that query. To achieve this, we introduced a joint upper and lower bound normalization (UL-normalization) framework and instantiated two versions of the UL-normalization, $V_1$ $V_2$ , for two popular IR evaluation metric $nDCG$ and  $MAP$, essentially creating four new evaluation metrics.

The next challenge in our work was to estimate a more realistic query-specific lower-bound for above two metric. For this estimation, we argued that a reasonable ranking method should be at least as good as a random ranking method, so a more realistic lower-bound should be the score expected by mere random ranking of the document collection rather than the current practice of assuming zero as lower-bound across all queries. Using probability and permutation theory, we  derived a closed-form formula to compute the expected $DCG$ in case of random ranking. The proof was completed by showing that expected relevance label of a document at position $i$ is actually independent of the position and can be replaced by the expected relevance label of the document collection associated with the particular query in the validation data-set. For expected $SP$ we also use probability and induction to prove the correctness of our assumption. The derivation details can be found in each case study section.

\smallskip
\noindent\textbf{Depth of Impact:} Using two publicly available web search and learning-to-rank data-sets, we conducted extensive experiments with eight popular LETOR methods to understand the implications $DCG^{U L}$ and $MSP^{U L}$. The implications are briefly summarized as below:

\begin{enumerate}[leftmargin=*,itemsep=0ex,partopsep=0.5ex,parsep=0ex]

\item Kendall's $\uptau$ rank correlation coefficient test on two different rankings of multiple LETOR methods, where the ranks are induced by both traditional metric (i.e.$nDCG$ and $MAP$) vs UL-normalized metrics(i.e. $DCG^{U L}$ and $MSP^{U L}$ ) yields \textbf{different conclusions} regarding the relative ranking of multiple LETOR methods.

\item Statistical Significance tests can lead to \textbf{conflicting conclusions} regarding the relative performance between a pair of LETOR methods, when comparing them in terms of traditional metrics vs UL-normalized metrics scores.

\item The above two observations are more prominent in case of \textit{Uninformative}
query collection.
\end{enumerate}

Next, we systematically compared the traditional  evaluation metric and  UL-normalized metrics from two important perspectives:  \textit{distinguishability} and \textit{consistency}. The findings are briefly summarized below.

\begin{enumerate}[leftmargin=*,itemsep=0.5ex,partopsep=0.5ex,parsep=0.5ex]

\item Discriminative power analysis and PAD scores suggest that our metric can better \textbf{distinguish} between two closely performing LETOR methods. These results were confirmed through Student's t-test and PAD score analysis.

\item For \textit{consistency}, $MSP^{U L}_{V_2}$ achieves the \textbf{lowest} swap rate across a data-sets comparison as well as the \textbf{lowest} swap rate while we compare the ranking results from \textit{uninformative} vs. \textit{ideal} query sets. On the other hand, the proposed $DCG^{U L}$ metric is identical to the original $nDCG$ metric in terms of \textbf{consistency} across different data-sets as well as across \textit{Uninformative}/ \textit{Ideal} query sets within the same data-set.  .

% kendall's $\tau$ correlation

\item All above experiments reveal that the impact of LB normalization is \textbf{more substantial} in case of ``{Uninformative}'' queries in comparison to ``Ideal'' queries, suggesting, LB normalization is crucial when the validation set contains a large number of \textit{Uninformative} queries (i.e., the ranking methods fail to perform significantly better than the randomly ranked output).

\end{enumerate}

\smallskip

\noindent\textbf{Breadth of Impact:} The proposed LB-normalization technique is very general and can be potentially extended to other IR evaluation metrics like ERR, which is an exciting future direction. Another direction can be to investigate such LB normalization for evaluation in domains other than IR, for example, ROUGE metric from the text summarization and NLP literature.

\smallskip
\noindent\textbf{Final Words:} The key take-away message from this paper is the following: \textit{{The IR community should consider lower-bound (LB) normalization seriously while evaluating any IR system.}} Our work takes a first step towards this important direction and can serve as a pilot study to demonstrate the importance and implications of LB normalization.